\documentclass[sigconf,screen]{acmart}

\AtBeginDocument{%
  }

\setcopyright{acmlicensed}
\acmDOI{10.1145/3650212.3652127}
\acmYear{2024}
\copyrightyear{2024}
\acmSubmissionID{issta24main-p491-p}
\acmISBN{979-8-4007-0612-7/24/09}
\acmConference[ISSTA '24]{Proceedings of the 33rd ACM SIGSOFT International Symposium on Software Testing and Analysis}{September 16--20, 2024}{Vienna, Austria}
\acmBooktitle{Proceedings of the 33rd ACM SIGSOFT International Symposium on Software Testing and Analysis (ISSTA '24), September 16--20, 2024, Vienna, Austria}
\received{16-DEC-2023}
\received[accepted]{2024-03-02}

\usepackage{algorithmic}
\usepackage{graphicx}
\usepackage{textcomp}
\usepackage{xcolor}
\usepackage{multirow}

\usepackage{booktabs}
\usepackage{enumitem}
\usepackage{soul}
\usepackage{ulem}
\usepackage{diagbox} 
\usepackage[most]{tcolorbox}  
\usepackage{color}
\usepackage{xcolor}
\usepackage{tikz}
\usepackage{tikz-qtree}
\usepackage{pgfplots}
\usepackage{varwidth}
\usepackage{flushend}
\usetikzlibrary{patterns}
\usepackage{url}
\usepackage{subfigure}
\usepackage{threeparttable}
\usepackage{colortbl}
\usepackage{xcolor}
\usepackage{balance}

\begin{document}

\title{Bridge and Hint: Extending Pre-trained Language Models for Long-Range Code}

\author{Yujia Chen}
\authornotemark[2]
\affiliation{%
  \institution{Harbin Institute of Technology}
  \city{Shenzhen}
  \country{China}
}
\email{yujiachen@stu.hit.edu.cn}

\author{Cuiyun Gao}
\authornote{Corresponding author.}
\affiliation{%
  \institution{Harbin Institute of Technology}
  \city{Shenzhen}
  \country{China}
}
\email{gaocuiyun@hit.edu.cn}

\author{Zezhou Yang}
\authornotemark[2]
\affiliation{%
  \institution{Harbin Institute of Technology}
  \city{Shenzhen}
  \country{China}
}
\email{yangzezhou@stu.hit.edu.cn}

\author{Hongyu Zhang}
\authornotemark[3]
\affiliation{%
  \institution{Chongqing University}
  \city{Chongqing}
  \country{China}
}\email{hyzhang@cqu.edu.cn}

\author{Qing Liao}
\authornotemark[2]
\affiliation{%
  \institution{Harbin Institute of Technology}
  \city{Shenzhen}
  \country{China}
}
\email{liaoqing@hit.edu.cn}

\newcommand{\tool}{EXPO}
\newcommand\etal{{\it{et al.\ }}}
\definecolor{light-gray}{gray}{0.9}    
\definecolor{mygray}{gray}{0.8}    
\definecolor{deep-gray}{gray}{0.7}    
\newcommand{\lgg}{\cellcolor{light-gray}}
\newcommand{\g}{\cellcolor{mygray}}
\newcommand{\hgg}{\cellcolor{deep-gray}}

\begin{abstract}
In the field of code intelligence, effectively modeling long-range code poses a significant challenge. Existing pre-trained language models (PLMs) such as UniXcoder have achieved remarkable success, but they still face difficulties with long code inputs. This is mainly due to their limited capacity to maintain contextual continuity and
memorize the
key information over long-range code.
To alleviate the difficulties, we propose
{\textbf{\tool}}, a framework for \textbf{EX}tending \textbf{P}re-trained language models for l\textbf{O}ng-range code.
{\tool} incorporates two innovative memory mechanisms we propose in this paper: \textit{Bridge Memory} and \textit{Hint Memory}. \textit{Bridge Memory} uses a tagging mechanism to connect disparate snippets of long-range code, helping the model maintain contextual coherence. \textit{Hint Memory} focuses on crucial code elements throughout the global context, such as package imports, by integrating a $k$NN attention layer to adaptively select
the relevant code elements.
This dual-memory approach bridges the gap between understanding local code snippets and maintaining global code coherence, thereby enhancing the model's overall comprehension of long code sequences.
We validate the effectiveness of {\tool} on five popular pre-trained language models such as UniXcoder and two code intelligence tasks including API recommendation and vulnerability detection. Experimental results demonstrate that {\tool} significantly improves the pre-training language models. 

\end{abstract}


\maketitle

\section{Introduction}
    Code intelligence is an important research
direction in the field of software engineering, aimed at using artificial intelligence technologies
to help software developers improve the development efficiency 
\cite{DBLP:conf/sigsoft/WangYGP0L22}. With the increasing scale of modern software development projects,
effectively modeling and understanding long-range code sequences has become a major challenge in the field of code intelligence. Pre-trained language models (PLMs), such as UniXcoder~\cite{DBLP:conf/acl/GuoLDW0022} and CodeT5~\cite{DBLP:conf/emnlp/0034WJH21}, have shown remarkable success in multiple code intelligence tasks, such as API recommendation~\cite{ptm4api,DBLP:conf/emnlp/KangW00Y21} and vulnerability detection~\cite{DBLP:journals/tosem/ZouZXLJY21,DBLP:conf/nips/ZhouLSD019,DBLP:conf/ndss/LiZXO0WDZ18}
, demonstrating the great potential of PLMs in understanding and generating source code. However, existing PLMs are limited to shorter code inputs, which directly affects the effectiveness of the models in practical applications. For example, the input length of the UniXcoder is limited to 512 tokens, while according to Guo et al.'s work~\cite{DBLP:conf/icml/GuoXD0M23}, 
the average length of Python source code files on GitHub is 2090 tokens after tokenization, and 24\% of the files have a length longer than 2,048 tokens, which means a large number of code files are beyond the processing capacity of existing PLMs. 
For the large language model, CodeGen~\cite{DBLP:conf/iclr/NijkampPHTWZSX23}, its maximum input length is only 2,048 tokens.
This highlights the need for models that can handle longer code sequences in order to be more practical and useful in the real-world. For long code sequences, PLMs face the following challenges:

\begin{figure*}
    \centering
    \includegraphics[scale=0.38]{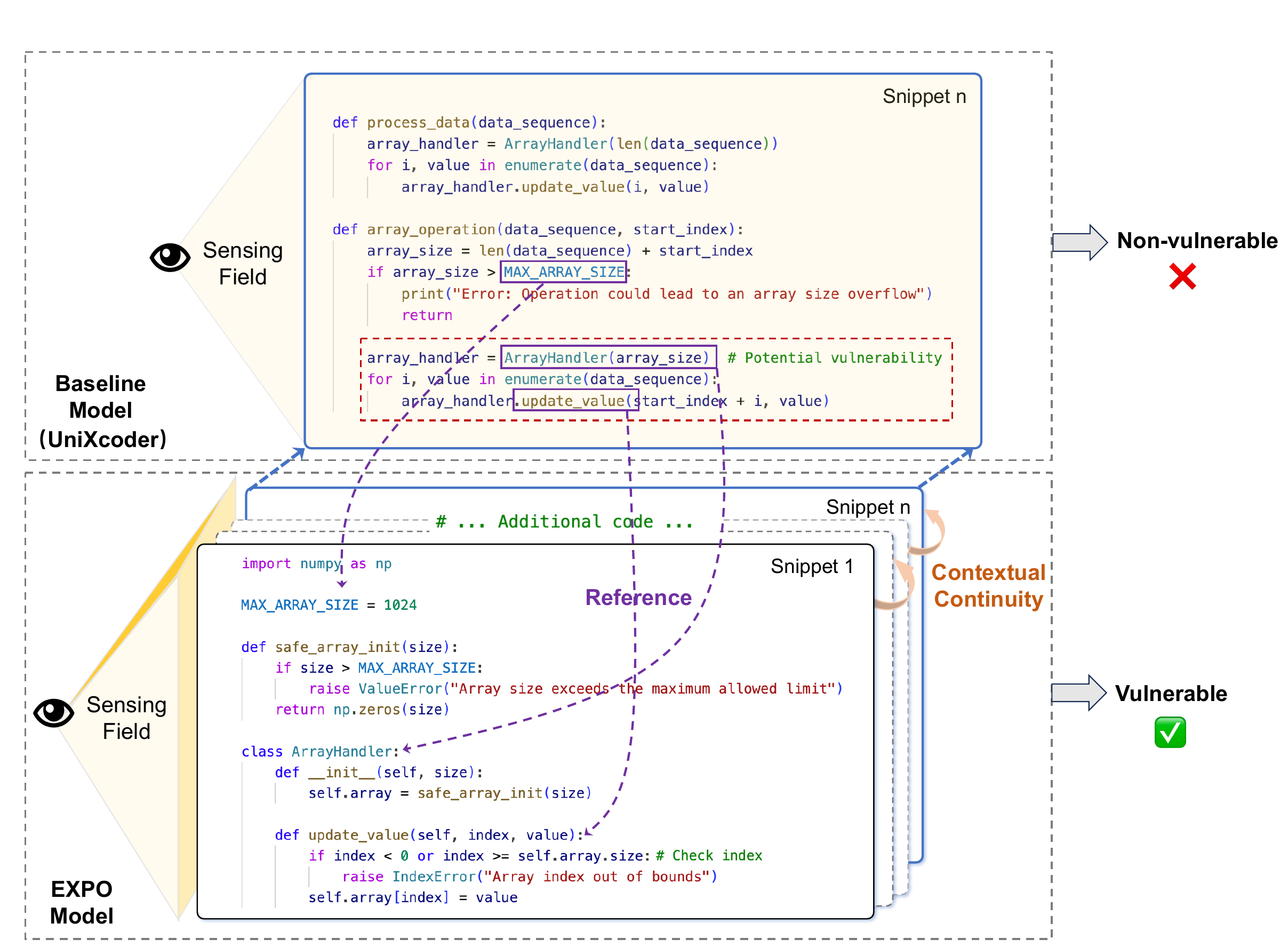}
    \caption{An example of vulnerability detection in a long-range code sequence.}
    \label{fig:motivation}
\end{figure*} 

\textit{1) Maintaining contextual continuity}. In long code sequences, capturing long-distance dependencies is the basis for understanding the entire program, which maintains the logical coherence between different parts of the code. For example, in code related to database operations, the initial connection setup might be at the beginning, while several related database queries and update operations might be spread in the following hundreds of lines. Therefore, models need to span extensive lines of code and accurately ``bridge'' these distributed elements to ensure a consistent understanding of the entire program's functionality. However, existing PLMs mainly focus on relatively short code sequences, resulting in performing poorly in maintaining the coherence of contextual information across long-range code~\cite{DBLP:conf/acl/GuanMFLDH20}.

\textit{2) Memorizing key information}. In source code, certain elements within
global scope, such as package imports, global variables and function definitions can be utilized
throughout the program. For example, a global configuration variable declared at the beginning of a program may be repeatedly referenced throughout the entire source code. Additionally, comments, though not related to the
program execution, provide crucial insights for understanding the intent and logic of the code~\cite{DBLP:conf/icse/CasalnuovoBDDM20}. This requires that PLMs not only identify these elements but also effectively memorize and utilize them throughout the code analysis process. However, existing PLMs often fail to accurately recognize 
these global code elements, leading to misinterpretations of the program's functionality.

These challenges highlight the limitations of existing PLMs in dealing with long code sequences. They struggle to maintain context information across extensive lines of code and lack in identifying and utilizing key code elements. Therefore, effective solutions to these issues are essential to advance the capabilities of PLMs to process long-range code.

\noindent \textbf{Our work.} In this paper, we propose {\textbf{\tool}}, a framework for \textbf{EX}tending \textbf{P}re-trained language models for l\textbf{O}ng-range code.
The core of {\tool} consists of two innovative memory mechanisms: \textit{Bridge Memory} and \textit{Hint Memory}, which work together to enhance the capabilities of PLMs in processing long code sequences. \textit{1) Bridge Memory}. For long code sequences, we first segment the code into fixed-length snippets that serve as inputs to the language model. At the beginning of each code snippet, we insert special bridge tokens that aggregate information from the current snippet. Subsequently, the context carried by these tokens is recurrently passed along sequences, which facilitates the flow and connection of information between snippets. This mechanism ensures continuity of information and integrity of context within long code sequences, where functions and variables may be defined and called in different parts of the code. \textit{2) Hint Memory}. In the code parsing phase, we identify and collect global declarations, such as package imports and class definitions, along with comments described in natural language, which are key clues for understanding the code. By storing the attention key-value pairs of these elements in a hint bank, our model can adaptively retrieve them through a $k$NN attention layer. This allows the model to access a rich global context when analyzing the current code snippet, thereby enhancing a comprehensive understanding of the code's logic and functionality. Together, {\it Bridge Memory} and {\it Hint Memory} enable {\tool} can not only thread
individual code snippets but also incorporate relevant
code structures across the entire long source code, aiming at mitigating
the limitations of PLMs in long-range code modeling.

We evaluate {\tool} on five state-of-the-art PLMs, including two encoder-only models, RoBERTa~\cite{DBLP:journals/corr/abs-1907-11692} and CodeBERT~\cite{DBLP:conf/emnlp/FengGTDFGS0LJZ20}, one decoder-only model, CodeGPT~\cite{DBLP:conf/nips/LuGRHSBCDJTLZSZ21} and two encoder-decoder models, CodeT5 and UniXcoder. We choose two common code intelligence tasks, including one generation task, API recommendation and one understanding task, vulnerability detection. Experimental results show that {\tool} improves the PLMs by 157.75\% $\sim$ 239.83\% on API recommendation in terms of average MRR and 5.60\% $\sim$ 46.36\% on vulnerability detection in terms of average F1 score. Besides, {\tool} achieves an average 29.42\% increase compared to the popular large language models (LLMs) such as ChatGPT on vulnerability detection regarding the F1 score metric.

\noindent \textbf{Contributions.}
In summary, our main contributions in this paper are as follows:

\begin{itemize}[leftmargin=*]
\item To the best of our knowledge, we are the first 
to propose a general framework to empower pre-trained models for effective long-range code modeling.
\item We propose a dual-memory mechanism, including \textit{Bridge Memory} for maintaining the coherence between code snippets, and \textit{Hint Memory} for capturing key code elements, together enhancing the modeling capability of long code sequences.
\item We conduct extensive experiments to evaluate {\tool} on two code intelligence tasks including API recommendation and vulnerability detection. The experimental results demonstrate that {\tool} substantially improves the performance of PLMs on both tasks.
\end{itemize}


\section{Motivating Example}  \label{sec:motivation}
    
In this section, we elaborate on the motivation for the framework design. 
Figure~\ref{fig:motivation} illustrates an example comparing the baseline model and the {\tool} model on vulnerability detection in long-range source code. As can be seen, the baseline model fails to identify it as non-vulnerable, whereas the {\tool} model successfully classifies it as vulnerable. The failure of the baseline model is related to the following two challenges:

\textbf{Challenge 1: Maintaining contextual continuity.} In this example, the baseline model's ``\textit{Sensing Field}'' only covers ``Snippet n'', which shows the {\tt ArrayHandler} class and its {\tt update\_value} method. This method updates array values but does not perform boundary checks on the array's index. However, in ``Snippet 1'', the {\tt update\_value} method imposes a requirement on the array index (the index needs to be greater than 0 and less than {\tt MAX\_ARRAY\_SIZE}). Due to the baseline model's narrow ``Sensing Field'', it fails to bridge to the functional implementation of the {\tt update\_value} method in ``Snippet 1'' and thus does not detect this potential vulnerability.

\textbf{Challenge 2: Memorizing key information.} The baseline model also does not \textit{reference} the global variable from ``Snippet 1'' - the {\tt MAX\_ARRAY\_SIZE} constant. This constant defines the boundary for array operations, which is crucial for avoiding buffer overflows. When ``Snippet n'' performs array updates through the {\tt update\_value} method, it should ensure the size constraint for safety. However, the baseline model, with its limited ``Sensing Field'', does not recognize the need to reference {\tt MAX\_ARRAY\_SIZE} during these operations, also resulting in a missed detection of the vulnerability.

Endowed with a broader ``\textit{Sensing Field}'', the {\tool} model is capable of analyzing
additional code snippets, providing it with a comprehensive view.
This allows the {\tool} model to capture the crucial reference to the global variable {\tt MAX\_ARRAY\_SIZE}.  
With this broader view, the model understands the necessity of performing boundary checks against {\tt MAX\_ARRAY\_SIZE} prior to invoking {\tt update\_value}. As a result, the {\tool} model could identify the code as ``vulnerable'' by recognizing the risk of a buffer overflow.

\section{Approach} \label{sec:approach}
    \begin{figure*}
    \centering
    \includegraphics[scale=0.41]{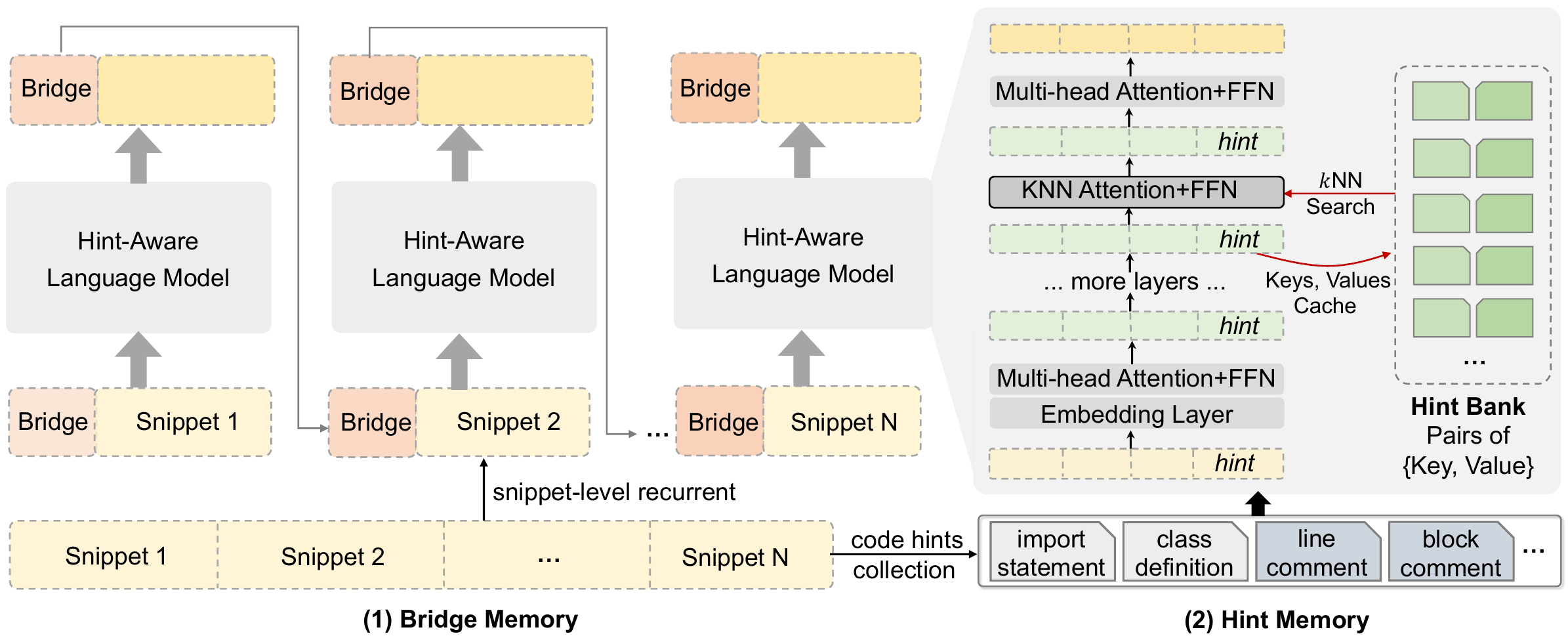}
    \caption{The overview of {\tool}.}
    \label{fig:framework}
\end{figure*}

In this section, we propose {\tool}, a framework to empower PLMs for effective
long-range code modeling. We first present the overview of {\tool} and then describe its details in the following subsections.

\subsection{Overview}

We provide an overview of {\tool} in Figure~\ref{fig:framework}, where {\tool} contains two major mechanisms, {\it Bridge Memory} and {\it Hint Memory}.
Our framework works as follows:
(1) For a long code sequence, we first segment it into some fixed-length snippets and then feed them into the model as input.
(2) In our {\it Bridge Memory} mechanism, a ``[Bridge]'' tag is added at the start of each snippet. 
This tag aggregates information from the previous snippet and updates itself with information from the current snippet, ensuring a continuous flow of context through the whole code sequence.
(3) In our \textit{Hint Memory} mechanism, we first carefully select key elements such as global declarations and comments as hints. 
These hints are stored as attention key-value pairs in a centralized hint bank.
A $k$NN attention layer then retrieves the relevant
pairs, allowing {\tool} could adaptively integrate global information into local snippet analysis.
In the end, {\tool} outputs a context-enriched
code representation, which can effectively
support downstream code intelligence tasks.


\subsection{Bridge Memory}
Given a long source code $\textbf{x} = (x_1, \cdots, x_C)$, where $C$ represents the total number of tokens, we first split the code into fixed-length snippets, each of length $L$. We denote the $i$-th snippet by $S_i = \{ x_{i,1}, \cdots, x_{i, L} \}$, where $0 \leq i \leq \lfloor C / L \rfloor$.
The \textit{Bridge Memory} mechanism begins by processing the first snippet $S_0$. The snippet is embedded using the Hint-aware Language Model (HLM) embedding layer, which yields the initial hidden state $H_{S_0}^{0}$. To facilitate the transfer of contextual information between different snippets, we introduce $m$ special bridge tokens to $S_0$, represented by the initial bridge hidden state $H_{bridge}^0$. The combined initial hidden state for $S_0$ at layer 0 is thus given by:
\begin{equation}
    \tilde{H}_{S_0}^0 = [ H_{bridge}^0 ; H_{S_0}^0 ]
\end{equation}
Subsequently, the Transformer layers within the HLM employ the hidden state from the previous layer $(n-1)$ to compute the hidden state for the current layer $n$:
\begin{equation}
\tilde{H}_{S_0}^n = f_{\theta_{HLM}^n} (\tilde{H}_{S_0}^{n-1}),
\end{equation}
where $f_{\theta_{HLM}^n}$ symbolizes the transformation function at layer $n$, and $n$ spans from 1 to $N$, with $N$ being the total number of layers in the HLM. This iterative process across the $N$ layers yields the final representation $\tilde{H}_{S_0}^N$, which is composed of the bridge token representation $H_{bridge}^N$ and the snippet representation $H_{S_0}^N$. The bridge token representation $H_{bridge}^N$  is obtained by attending to all current snippet tokens and updating its representation accordingly. It serves a dual purpose: aggregating contextual information for the current snippet $S_0$ and acting as the bridging component for the subsequent snippet $S_1$. Each subsequent snippet is processed similarly. Finally, this recurrent process outputs a context-enriched representation $H_{S_{\lfloor C / L \rfloor}  }^N$ of the long-range code.
Through the \textit{Bridge Memory} mechanism, snippet-level information is effectively propagated across the entire code sequence, thereby preserving contextual continuity — an important aspect for modeling, as illustrated in Figure~\ref{fig:framework}(1). 


\subsection{Hint Memory}

The {\it Hint Memory} mechanism focuses on the critical code elements in long source code. It consists of two parts: collecting code hints 
and utilizing these hints in the hint-aware language model as illustrated in Figure~\ref{fig:framework}(2). 


\subsubsection{Code Hint Collection} \label{subsec:hintCollection}

\begin{table}[tb]
\small
\caption{The {collected}
code hints{, with examples} 
extracted from the source code in Figure~\ref{fig:motivation}.}
\begin{tabular}{c|c|l}
\toprule
\textbf{Hint}      & \textbf{Description}     &\multicolumn{1}{c}{\textbf{Example}}      \\ \midrule
\begin{tabular}[c]{@{}c@{}}Import \\ Statement\end{tabular}    & \begin{tabular}[c]{@{}c@{}}Modules or libraries \\ to include\end{tabular}     & \verb|import numpy as np|                                                           \\ \midrule
\begin{tabular}[c]{@{}c@{}}Class \\ Definition\end{tabular}    & \begin{tabular}[c]{@{}c@{}}Description of a class \\ with methods\end{tabular} & \verb|class ArrayHandler|                                                            \\ \midrule
\begin{tabular}[c]{@{}c@{}}Function \\ Definition\end{tabular} & \begin{tabular}[c]{@{}c@{}}Function name, and \\ parameters\end{tabular}       & \verb|def __init__|                                                             \\ \midrule
\begin{tabular}[c]{@{}c@{}}Field \\ Definition\end{tabular}    & \begin{tabular}[c]{@{}c@{}}Variables in a class \\ or global\end{tabular}     & \begin{tabular}[c]{@{}l@{}}\verb|MAX_ARRAY_SIZE=1024|\end{tabular} \\ \midrule
\begin{tabular}[c]{@{}c@{}}Line \\ Comment\end{tabular}        & \begin{tabular}[c]{@{}c@{}}Description for \\ code statement\end{tabular}      & \verb|# Check index|                                                                               \\ \midrule
\begin{tabular}[c]{@{}c@{}}Block \\ Comment\end{tabular}       & \begin{tabular}[c]{@{}c@{}}Description for \\ code segment\end{tabular}        & \verb|/* ... */|     \\ 
\bottomrule
\end{tabular}
\label{tab:code_hints}
\end{table}


Code elements such as global declarations and comments are key clues for understanding source code. For example, in Figue~\ref{fig:motivation}, to call the {\tt update\_value} method of the {\tt Array Handler} class, the model needs to access the original definition of the class {\tt Array Handler}. We identify six types of code elements as code hints, which are detailed in Table~\ref{tab:code_hints}. Import statement, modules or libraries to include, clearly indicates which external modules or libraries the current code file depends on. This helps developers quickly understand the context of the code~\cite{DBLP:journals/ese/KulaGOII18}. Class definition, description of a class with methods, is crucial for understanding the structure and behavior patterns of the program, as the relationships between classes (such as inheritance and composition) define the architecture of the program~\cite{DBLP:conf/icse/Butler12}. Function definition, function name and parameters, provides the function's purpose, inputs, and expected outputs, which is crucial for understanding the program's behavior and how modules interact with each other~\cite{DBLP:conf/icse/McIntoshANKH11}. Field definition, variables in a class or global, is crucial for understanding the behavior of the program, especially when debugging or extending the code~\cite{DBLP:journals/ese/McIntoshKAH16}. Line comment, description for code statement, provides immediate explanations of code statements, helping developers quickly understand complex or non-obvious code logic~\cite{DBLP:conf/icse/BacchelliB13}. Block comment, description for code segment, provides a high-level overview of code segments, helping developers understand the purpose of the code segment and how it contributes to the overall functionality of the program~\cite{DBLP:conf/icse/PadioleauTZ09}. To collect these hints from source code, we use the tree-sitter~\cite{tree-sitter} tool to parse the code into an Abstract Syntax Tree (AST). 
In AST, we collect these important code hints and use a mask vector $B$ to represent these hints, i.e., $B_i$ = 1, if the $i$-th token is a code hint; otherwise $B_i$ = 0.

\subsubsection{Hint-Aware Language Model}
For an input snippet $S_i$, the Hint-Aware Language Model (HLM) performs a forward pass to encode the input $S_i$ into embedding space
and outputs the final layer's hidden states, $\tilde{H}_{S_i}^N$.
During the forward pass with the HLM for $S_i$, the key-value pairs used for multi-head self-attention of these code hints at the $n$-th Transformer layer are stored in the hint bank.
{\bf Hint Bank} is a cached head-wise vector, which contains attention key-value pairs of all the code hints in the previous inputs $ \{\tilde{K}, \tilde{V} \}$. 
Our HLM employs the hidden state of snippet $S_i$ at the $n$-th layer, $\tilde{H}_{S_i}^n$, along with cached pairs, to enhance its contextual representation through a special $k$NN attention layer.
For input $S_i$, the {\bf $\mathbf{k}$NN Attention Layer} first retrieves top-$K$ relevant cached key-value pairs via $k$NN search from the hint bank \(\{ \tilde{K}_{ij} ,\tilde{V}_{ij}\}_{j=1}^K \).
Finally, it uses a gated mechanism to fuse the cached attention pairs and local hidden state $\tilde{H}_{S_i}^n$.

\textit{1) Cached Hint Bank}. For the input snippet $S_i$, we cache the key-value pairs from the final layer's multi-head attention of the Transformer into the hint bank. 
This cached information is then used when processing the subsequent snippet $S_{i+1}$, aiming at enhancing the model's ability to understand and utilize global information in long source code. The code hints' key-value pairs are added to the hint bank using the formula below:
\begin{equation}
(\tilde{K}, \tilde{V}) = (K, V) \odot B
\end{equation}

\textit{ 2) $k$NN attention layer.} For the input snippet $S_i$, the model first retrieves the top-$K$ most relevant pairs of keys and values (\(\{ \tilde{K}_{ij} ,\tilde{V}_{ij}\}_{j=1}^K\)) from the hint bank via the $k$NN search. 
Next, the $k$NN attention layer employs a long-term memory fusion process to enable each snippet to attend to both local contexts and retrieved global contexts (i.e., code hints).
With the hidden state from the previous layer ($\tilde{H}^{n - 1}$) and the retrieved attention key-value pairs (\(\{ \tilde{K}_{ij},\tilde{V}_{ij}\}_{j=1}^K\)), the output hidden state for the $n$-th
{\it Hint Memory} layer $\tilde{H}_{S_i}^n$ is computed as:
\begin{align}
A_{local} &= \text{softmax}\left(\frac{QK^T}{\sqrt{d}}\right)V \\
A_{knn} &= \text{Concat}\{\text{softmax}\left( \frac{Q\tilde{K}_i^T}{\sqrt{d}}\right)\tilde{V}_i\}_{i=1}^K\\
\tilde{H}_{S_i}^n &= \text{sigmoid}(g) \cdot A_{local} + (1 - \text{sigmoid}(g)) \cdot A_{knn},
\end{align}
where $K$ is the number of retrieved attention key-value pairs in the hint cache for each snippet, and $g$ is a trainable head-wise gating vector. 
The hidden state output from
the previous layer $\tilde{H}^{n - 1}$ is linearly projected into attention queries, keys, and values $Q,K,V$ separately
via three matrices $W^Q,W^K,W^V$. It is worth noting that the retrieved attention key-value pairs in the hint bank are distinct to each snippet. 
The $k$NN attention layer in {\tool} fuses code hints (i.e., global information) with the local context, which makes {\tool} able to handle long source code better.

\section{Experimental Setup} \label{sec:setup}
    \subsection{Research Questions} \label{sec:RQ}

We conduct extensive experiments to evaluate the proposed approach with the aim of answering the following research questions:

\begin{itemize}[leftmargin=*]

\item {\textbf{RQ1:}} How effective is {\tool} in enhancing the ability of PLMs for long-range code modeling?
\item {\textbf{RQ2:}} What are the impacts of the two main mechanisms (i.e., \textit{Bridge Memory} and \textit{Hint Memory}) on {\tool}?
\item {\textbf{RQ3:}} What is
the performance of {\tool} compared with that of LLMs?
\item {\textbf{RQ4:}} How does {\tool} perform under different parameter settings?

\end{itemize}

\subsection{Baselines} \label{sec:baseline}

We apply {\tool} for five state-of-the-art pre-trained models, including two encoder-only models, RoBERTa~\cite{DBLP:journals/corr/abs-1907-11692} and CodeBERT~\cite{DBLP:conf/emnlp/FengGTDFGS0LJZ20}, one decoder-only model, CodeGPT~\cite{DBLP:conf/nips/LuGRHSBCDJTLZSZ21} and two encoder-decoder models, CodeT5~\cite{DBLP:conf/emnlp/0034WJH21} and UniXcoder~\cite{DBLP:conf/acl/GuoLDW0022}. RoBERTa is an improved version of the BERT~\cite{DBLP:conf/naacl/DevlinCLT19} model. CodeBERT is pre-trained on NL-PL pairs in six programming languages that achieve promising results on code intelligence tasks. CodeGPT is pre-trained by generating code from left to right in an auto-regressive manner on large amounts of code. CodeT5 adapts the T5~\cite{DBLP:journals/jmlr/RaffelSRLNMZLL20} model that considers the crucial token type information from identifiers. UniXcoder is a unified pre-trained model that incorporates semantic and syntax information from code comments and Abstract Syntax Tree (AST), which achieves state-of-the-art performance on various code intelligence tasks.

\subsection{Evaluation Tasks} \label{sec:task}

We conduct extensive experiments on two common code intelligence tasks: API recommendation and vulnerability detection.

\subsubsection{API recommendation:} API recommendation aims to automatically suggest appropriate Application Programming Interface (API) calls for specific programming tasks within a given code snippet~\cite{DBLP:journals/tse/ChenGRP0L23}. Recent work mainly formulates it as a sequence-to-sequence neural machine translation (NMT) task and involves pre-trained techniques to achieve better performance~\cite{DBLP:conf/icse/KimZT021,ptm4api}.

\begin{table}[tb]
\centering
\renewcommand\arraystretch{1.2}
\caption{Statistics of the datasets used in this paper. ``AR'' and ``VD'' denote the API recommendation task and vulnerability detection task, respectively. ``\#Files'' denotes the number of code files in the dataset. ``Avg.'' denotes the average code length in the dataset.}
\label{tab:datasets}
\scalebox{0.95}{\begin{tabular}{cc|cc|cc|cc}
\toprule
\multicolumn{2}{c|}{\multirow{2}{*}{\textbf{Task}}} & \multicolumn{2}{c|}{\textbf{Train}}  & \multicolumn{2}{c|}{\textbf{Valid}} & \multicolumn{2}{c}{\textbf{Test}} \\
\multicolumn{2}{c|}{}  &\textbf{\#Files} & \textbf{Avg.}  &\textbf{\#Files} & \textbf{Avg.}  &\textbf{\#Files} &\textbf{Avg.}   \\
\midrule
\multicolumn{1}{c|}{\multirow{2}{*}{AR}}  & Java & 143,504  &  1,689  & 17,938  &  1,226  & 17,938  & 2,002  \\
\multicolumn{1}{c|}{}  & Python & 26,910  & 2,394   &  3,364 &  1,309  & 3,364 & 2,766 \\
\midrule
\multicolumn{1}{c|}{\multirow{2}{*}{VD}}  & Java & 500  &  5,485  &  -  &  -  & 56  & 5,296  \\
\multicolumn{1}{c|}{}  & Python & 490  &  6,975  &  - &  -  & 54  & 6,143 \\
\bottomrule 
\end{tabular}
}
\end{table}

\textbf{Dataset.} To evaluate the performance of API recommendations, we use the APIBench-C benchmark dataset collected by Peng et al.~\cite{DBLP:journals/tse/PengLGLWGL23}. It contains complete source code files covering Java and Python projects downloaded from GitHub~\cite{Github} in different domains. In this study, considering our time and resource limitations, we use datasets from the ``General'' domain, consisting of 1,056,790 Java files and 230,064 Python files. Considering the evaluation is mainly for long-range code, 
we eliminate code files with fewer than 512 code tokens after tokenization, as the input length of existing PLMs is limited to 512 tokens. We also exclude excessively long files with more than 10,000 code tokens. After this filtering, we obtain 143,504 Java files and 26,910 Python files, which are randomly partitioned into training, validation, and test sets with a ratio of 8:1:1, respectively. The specific data statistics are shown in Table~\ref{tab:datasets}.

\textbf{Metrics.} We adopt two widely-used metrics~\cite{DBLP:journals/tse/ChenGRP0L23, DBLP:conf/icse/WeiHH0022, DBLP:journals/tse/PengLGLWGL23} for API recommendation evaluation: SuccessRate and Mean Reciprocal Rank (MRR). 

\begin{itemize}[leftmargin=*]

\item SuccessRate@k (SR@k) evaluates the ability of a model to recommend correct APIs based on the top-$k$ returned results regardless of the orders. We set the value of $k$ to 1, 3 and 5, respectively.

\begin{equation}
    SR@k = \frac{\sum_{i=1}^N \text{HasCorrect}_k(q_i)} {N},
\end{equation}

where $\text{HasCorrect}_k(q)$ returns 1 if the top-$k$ results of query $q$ contain the correct API, otherwise it returns 0. $N$ denotes the number of samples.

\item MRR measures the average ranking of the first correct API in the ranking list.

\begin{equation}
     MRR = \frac{\sum_{i=1}^N 1/ \text{firstpos} (q_i)} {N},
\end{equation}

where $\text{firstpos}(q)$ returns the position of the first correct API in the results, if it cannot find a correct API in the results, it returns $+\infty$.
\end{itemize}

\subsubsection{Vulnerability Detection:} Vulnerability detection aims to identify whether a given source code contains vulnerabilities that may pose risks to software systems, such as resource leakage. Recent work mainly formulates it as a binary classification task~\cite{DBLP:conf/emnlp/FengGTDFGS0LJZ20,DBLP:conf/acl/GuoLDW0022,DBLP:conf/emnlp/0034WJH21}.

\textbf{Dataset.} To evaluate the performance of vulnerability detection, we use the CrossVul~\cite{DBLP:conf/sigsoft/NikitopoulosDLM21} dataset, which comprises 556 Java files and 544 Python files collected from GitHub.
Due to the limited learning data, we do not partition a separate validation set from the files. The dataset is split into the training set and test set in the proportion of 9:1. We employ ten-fold cross-validation instead of a static validation set to validate the model's performance, ensuring its robustness and parameter optimization. Ten-fold cross-validation involves first dividing the training dataset into ten equal parts, and then using nine parts for training
and the remaining one part for validation iteratively.

\textbf{Metrics.} For vulnerability detection,
we follow the previous work~\cite{DBLP:conf/sigsoft/NikitopoulosDLM21} to evaluate the results by the Accuracy and F1 score metric. 
As a binary classification problem, a true positive occurs when the model correctly detects a true vulnerability. In contrast, a false positive is when the model
detects a vulnerability that is not exploitable. True and false negatives are defined analogously.

\begin{itemize}[leftmargin=*]

\item Accuracy measures the abilities of
the 
model to make a correct prediction, i.e., whether a code snippet is vulnerable or not.

\begin{equation}
Acc. = \frac{ \text{\# True Positives} + \text{\# True Negatives}}{N}
\end{equation}

\item F1 score is the harmonic mean of precision and recall, providing a balance between them. It is especially useful in situations where there is an uneven class distribution, as is often the case in vulnerability detection:

\begin{equation}
F1 = 2 \times \frac{ \text{Precision} \times \text{Recall}}{\text{Precision} + \text{Recall}}
\end{equation}
where Precision is $\frac{\text{\# True Positives}}{\text{\# True Positives} + \text{\# False Positives}}$ and Recall is \\ $\frac{\text{\# True Positives}}{\text{\# True Positives} + \text{\# False Negatives}}$.

\end{itemize}

\subsection{Implementation Details}

All baseline PLMs are downloaded from the HuggingFace Hub~\cite{HuggingFace}. We use {\tool} to fine-tune these PLMs for two tasks: API recommendation on the APIBench-C dataset and vulnerability detection on the CrossVul dataset. We set the maximum code length $C$ to 4096 and the snippet length $L$ to 512. Within the 12 layers of each PLM, the 11-th layer is considered as the $k$NN attention layer. For the hyper-parameters, $m$ in \textit{Bridge Memory} and $K$ in \textit{Hint Memory}, we set them to one and 32 by default, respectively. The impact of these parameters is further discussed in Section~\ref{sec:para}. We fine-tune all baseline PLMs over 10 epochs for each task. During fine-tuning, we employ the Adam optimizer~\cite{DBLP:journals/corr/KingmaB14} with a batch size of 8 and a learning rate of 2e-5. All experiments are conducted on three servers with 12 NVIDIA A100-40GB GPUs each, and a server with 3 NVIDIA V100 GPUs.


\section{Evaluation} \label{sec:results}
    

\subsection{Effectiveness of {\tool} (RQ1)}

\begin{table*}[t]
\centering
\renewcommand\arraystretch{0.95}
\caption{Performance comparison between the base and {\tool} 
models.}
\label{tab:comparison_main}

\begin{tabular}{cc|ccccc|ccccc|ccc|ccc}
    \toprule

    \multicolumn{2}{c|}{\multirow{3}{*}{\textbf{Model}}}& & \multicolumn{8}{c}{\textbf{API Recommendation}}  & & & \multicolumn{4}{c}{\textbf{Vulnerability Detection}}&    \\
    \cmidrule{4-11}
    \cmidrule{14-17}
    \multicolumn{2}{c|}{} &  & \multicolumn{4}{c|}{\textbf{Java}} & \multicolumn{4}{c}{\textbf{Python}}& & &  \multicolumn{2}{c|}{\textbf{Java}} & \multicolumn{2}{c}{\textbf{Python}} &\\
    \multicolumn{2}{c|}{} & & SR@1  & SR@3  & SR@5 & MRR  & SR@1  & SR@3  & SR@5 & MRR & & & Acc.   & F1  & Acc.   & F1& \\
    \midrule
    \multicolumn{1}{c|}{\multirow{2}{*}{RoBERTa}}  & Base &  &  21.13  & 29.23   &32.41    & 0.234 &  15.66    & 25.47  & 29.36   &  0.202 & & &  55.36  &  54.17 & 52.70  & 43.44 &  \\
    \multicolumn{1}{c|}{}  & {\tool} &  &\textbf{54.33}    & \textbf{61.90}  &\textbf{63.92}   & \textbf{0.582}  &  \textbf{51.19}    & \textbf{63.20}    & \textbf{68.49}   & \textbf{0.559} & & & \textbf{57.14}  &  \textbf{71.43} & \textbf{53.07} & \textbf{69.88} & \\
    \midrule
    \multicolumn{1}{c|}{\multirow{2}{*}{CodeBERT}}  & Base &   & 22.18    & 31.32  & 34.68   &0.270   &16.56  & 26.10  & 29.54   &0.213  & & &  53.57   &  51.34   &  50.00 &  45.82 &  \\
    \multicolumn{1}{c|}{}   & {\tool} &  &\textbf{59.23}     &\textbf{65.83}   &\textbf{67.66}    & \textbf{0.625}   & \textbf{55.92}    & \textbf{64.86}  & \textbf{68.07}  &  \textbf{0.605} && & \textbf{55.35}  & \textbf{61.54}   & \textbf{55.56}  &\textbf{64.71}    &  \\
    \midrule
    \multicolumn{1}{c|}{\multirow{2}{*}{CodeGPT}}  & Base  &   & 13.70    & 16.91   & 19.93   &0.155  & 17.90  & 19.69  & 21.02   & 0.188 & && 51.79    & 68.23    & 53.70  & 65.87  & \\
    \multicolumn{1}{c|}{}     & {\tool} &  &  \textbf{48.27}    &
\textbf{52.14}   & \textbf{55.14}    &  \textbf{0.503} & \textbf{48.37}    & \textbf{52.35}    &\textbf{55.26}    & \textbf{0.504} & & & \textbf{58.93} & \textbf{70.89} &\textbf{55.56}  & \textbf{70.73}   & \\
    \midrule
    \multicolumn{1}{c|}{\multirow{2}{*}{CodeT5}}  & Base   &   & 22.66    & 33.39  &  37.75  & 0.271   & 12.54    & 19.02    & 21.22   &0.159 & & &53.70   & 53.57   & 46.30  & 46.28 &  \\
    \multicolumn{1}{c|}{}    & {\tool}  & & \textbf{64.03}    & \textbf{73.01}  & \textbf{76.39}   & \textbf{0.688}   & \textbf{63.08}   & \textbf{71.88}   & \textbf{74.73}    & \textbf{0.677} && & \textbf{55.36}   & \textbf{66.67} & \textbf{55.56}  & \textbf{68.42}   & \\
    \midrule
    \multicolumn{1}{c|}{\multirow{2}{*}{UniXcoder}}  & Base  &   &  25.29 & 39.46   &44.06    & 0.299   & 17.21   & 26.25    &  30.41  & 0.222 && & 50.00  &  60.16   & 50.00  & 64.94  &\\
    \multicolumn{1}{c|}{}      & {\tool} & & \textbf{53.55}   & \textbf{70.80} &  \textbf{81.38} &  \textbf{0.671}   & \textbf{62.04 } & \textbf{75.18}  & \textbf{83.02}   & \textbf{0.670} && & \textbf{60.71}   & \textbf{66.67}  & \textbf{62.96}  &\textbf{68.75}  & \\
    \bottomrule
    
    \end{tabular}

\end{table*}

\noindent \textbf{Experimental Design.} To answer this research question, we use {\tool} for the five PLMs listed in Section~\ref{sec:baseline} on two downstream code intelligence tasks including API recommendation and vulnerability detection, described in Section~\ref{sec:task}.

\noindent \textbf{Results.} Table~\ref{tab:comparison_main} presents the performance comparison between the {\tool}-extended models and corresponding
original baseline (base) models for the two downstream tasks. Using the Wilcoxon signed-rank test~\cite{wilcoxon1992individual} (with a \textit{p} value $<$ 0.001), we validate that the performance improvements of {\tool} are statistically significant.

\noindent \textbf{Analyses.} As shown in Table~\ref{tab:comparison_main}, {\tool} can consistently achieve the best performance on all metrics and tasks. 
1) For the encoder-only model CodeBERT, {\tool} outperforms the base model by 167.04\% on SR@1, 110.18\% on SR@3, 95.10\% on SR@5 and 131.48\% on MRR for API recommendation (Java). 
2) For the decoder-only model CodeGPT, on API recommendation, {\tool} achieves an improvement of 224.52\% and 168.08\% in terms of MRR in Java and Python, respectively. On vulnerability detection, compared with base, {\tool} also improves it by 13.78\% and 3.46\% regarding the Accuracy metric in Java and Python, respectively.
3) For the encoder-decoder model UniXcoder on API recommendation, the average SR@1 and MRR of {\tool} are 57.79 and 0.670, showing improvements of 171.95\% and 157.19\% over the base model, respectively. Furthermore, {\tool} achieves the highest SR@5, with a score of 81.38 and 83.02 for Java and Python, respectively. As for vulnerability detection in Java, we observe that {\tool} outperforms the base by 21.42\% and 10.82\% in terms of Accuracy and F1 score, respectively. For the following research questions, considering that UniXcoder demonstrates comparable performance to the other
PLMs, we use UniXcoder as the base model for analysis for saving computation resources.

Overall, {\tool} can be effectively employed to improve the performance of various types of PLMs for long-range code input, with particularly notable improvements of 157.75\% $\sim$ 239.83\% on API recommendation in terms of average MRR and 5.60\% $\sim$ 46.36\% on vulnerability detection in terms of average F1 score.
Moreover, we observe that {\tool} has the most significant impact on CodeGPT and CodeT5. For example, on API recommendation, the SR@k (k=1,3,5) of CodeT5 is improved by 153.87\% $\sim$ 403.03\% after employing {\tool}, while for CodeGPT, it is 162.89\% $\sim$ 252.34\%. Additionally, {\tool} is more helpful for API recommendation. Across all baseline PLMs, {\tool} improves SR@1 by at least 170.22\% for API recommendation in Python, and improves F1 score by at least 5.80\% for vulnerability detection in Python.

\begin{tcolorbox}[breakable,width=\linewidth-2pt,boxrule=0pt,top=3pt, bottom=3pt, left=4pt,right=4pt, colback=gray!15,colframe=gray!15]
\textbf{Answer to RQ1:} 
{\tool} can effectively empower various PLMs to model long-range source code, with particularly notable improvements of 157.75\% $\sim$ 239.83\% on API recommendation in terms of average MRR and 5.60\% $\sim$ 46.36\% on vulnerability detection in terms of average F1 score.
\end{tcolorbox}

\subsection{Impacts of Different Mechanisms in {\tool} (RQ2)}

\begin{table*}
\centering
\renewcommand\arraystretch{0.9}
\caption{Ablation study. ``w/o bridge'' and ``w/o hint'' denote removing \textit{Brideg Memory} and \textit{Hint Memory}, respectively. }
\label{tab:comparison_aba}

\begin{tabular}{cc|ccccc|ccccc|ccc|cc}
    \toprule
    \multicolumn{2}{c|}{\multirow{3}{*}{\textbf{Model}}} & \multicolumn{10}{c|}{\textbf{API Recommendation}} & \multicolumn{5}{c}{\textbf{Vulnerability Detection}}    \\
    \cmidrule{4-11} 
    \cmidrule{14-17} 
    \multicolumn{2}{c|}{}   &  & \multicolumn{4}{c|}{\textbf{Java}} & \multicolumn{4}{c}{\textbf{Python}}  &  & & \multicolumn{2}{c|}{\textbf{Java}} & \multicolumn{2}{c}{\textbf{Python}}  \\
    \multicolumn{2}{c|}{} & & SR@1  & SR@3  & SR@5 & MRR   & SR@1  & SR@3  & SR@5 & MRR &  & & Acc.   & F1 & Acc.   & F1   \\
    \midrule
    \multicolumn{1}{c|}{\multirow{3}{*}{RoBERTa}}  & {\tool}  &   & \textbf{54.33}    & \textbf{61.90}  &\textbf{63.92}   & \textbf{0.582}  & \textbf{51.19}    & \textbf{63.20}    & \textbf{68.49}   & \textbf{0.559} &  & & \textbf{57.14}    & \textbf{71.43}  &\textbf{53.70} & \textbf{69.88}    \\
    \multicolumn{1}{c|}{}    & w/o bridge & & 53.08    & 59.69   & 61.98   & 0.553   &49.73   &61.17    &65.96   & 0.547   &  & &51.78   & 67.46    & 52.40 & 61.33 \\
    \multicolumn{1}{c|}{}   & w/o hint  & & 51.66    & 58.32  & 60.89   & 0.547    &47.50     & 59.80    &64.59    &0.524  &  &   & 55.35  &  69.13  & 51.70  & 59.87  \\
    \midrule
    \multicolumn{1}{c|}{\multirow{4}{*}{CodeBERT}}   & {\tool}  &  &\textbf{59.23}     &\textbf{65.83}   &\textbf{67.66}   & \textbf{0.625}     & \textbf{55.92}    &\textbf{64.86}     &\textbf{68.07}  & \textbf{0.605} &  & & \textbf{55.35}    &\textbf{61.54} & \textbf{55.56}  &\textbf{64.71}    \\
    \multicolumn{1}{c|}{}  & w/o bridge & &57.37  & 64.47  & 66.19   &  0.608    &53.59   &63.10    &66.58   & 0.585   &  & &50.00   & 54.00 & 53.70  & 58.35  \\
    \multicolumn{1}{c|}{}  & w/o hint &  & 54.39    & 62.46  &  65.04  & 0.587   & 51.26    &  61.40   & 64.46   & 0.563  &  & & 52.35 & 52.48  & 52.35  & 54.54  \\
    \midrule
    \multicolumn{1}{c|}{\multirow{4}{*}{CodeGPT}}  & {\tool}  &  & \textbf{48.27}    & \textbf{52.14}   & \textbf{55.14}    & \textbf{0.503}      & \textbf{48.37}    & \textbf{52.35}    & \textbf{55.26}    & \textbf{0.504} &  & & \textbf{58.93}    & \textbf{70.89} & \textbf{55.56}  & \textbf{70.73}   \\
    \multicolumn{1}{c|}{}    & w/o bridge &  & 41.84    & 47.18  & 51.99   & 0.447    & 46.08  & 48.41  & 53.16    &0.481   &  & &55.35   & 69.13 & 53.70  & 68.87  \\
    \multicolumn{1}{c|}{} & w/o hint &  & 40.52    &  45.91  &  49.83    & 0.439  & 45.80  & 47.87  & 50.59  & 0.469  &  &    & 53.57   & 69.76   & 53.70  & 66.66 \\
    \midrule
    \multicolumn{1}{c|}{\multirow{4}{*}{CodeT5}}  & {\tool}  &   & \textbf{64.03}    & \textbf{73.01}  & \textbf{76.39}   & \textbf{0.688}    & \textbf{63.08}   & \textbf{71.88}   &\textbf{74.73}    & \textbf{0.677}  &&  &  \textbf{55.36}   & \textbf{66.67} & \textbf{55.56}  & \textbf{68.42} \\
    \multicolumn{1}{c|}{}  & w/o bridge & & 55.02   & 66.78  & 71.15   & 0.610  &   59.60  & 69.38    & 71.90  &0.645  & &  & 53.57  &  58.29  & 53.70   &  61.53 \\
    \multicolumn{1}{c|}{}  & w/o hint  & & 52.84  & 65.98  &  68.85  &  0.595    & 56.65   &  65.27   & 69.58  & 0.610 & &   & 50.00   &61.11    & 50.00   & 66.67 \\
    \midrule
    \multicolumn{1}{c|}{\multirow{4}{*}{UniXcoder}}  & {\tool} & & \textbf{53.55}   & \textbf{70.80} &  \textbf{81.38} & \textbf{0.671}  & \textbf{62.04}  & \textbf{75.18}  & \textbf{83.02}   &  \textbf{0.669}   &  & & \textbf{60.71}   & \textbf{66.67}  & \textbf{62.96}  &\textbf{68.75}      \\
    \multicolumn{1}{c|}{}  & w/o bridge & & 51.85   & 67.14  & 73.35   & 0.632  &  61.80   & 73.92   &80.64   &0.658  &   &  & 51.78  & 65.66  & 53.70  & 61.33    \\
    \multicolumn{1}{c|}{} & w/o hint &  & 50.27    & 65.51  & 72.95 & 0.611  &    58.55   &  68.31     & 73.65  & 0.530   & & & 48.21  & 56.71   & 53.70  & 59.13 \\
    \bottomrule
    \end{tabular}

\end{table*}

\noindent \textbf{Experimental Design.} For
this research question, we perform ablation studies by considering the following two variants of {\tool}.

\begin{itemize}[leftmargin=*]

\item {$\text{\tool}_{\text{w/o\ bridge}}$:} In this variant, we remove the \textit{Bridge Memory}. To ensure a fair comparison, we directly aggregate the representations of each snippet, allowing the model to perceive the whole context.
 
\item {$\text{\tool}_{\text{w/o\ hint}}$:} In this variant, we exclude \textit{Hint Memory} and rely solely on the \textit{Bridge Memory} to maintain context continuity of long-range code.

\end{itemize}

\noindent \textbf{Results.} Table~\ref{tab:comparison_aba} shows the results of {\tool} compared with the two variants on API recommendation and vulnerability detection. Both variants have performance degradation.

\noindent \textbf{Analyses.}
Experimental results reveal that removing the \textit{Hint Memory} mechanism leads to a large drop in {\tool}'s performance. 
For example, for API recommendation, the average SR@k (k=1,3,5) and MRR of UniXcoder decrease by 9.14\% and 18.02\%, respectively, while the performance on vulnerability detection experiences a drop of 21.59\% and 16.92\% in average Accuracy and F1 score, respectively. 
This is because without memorizing key information such as package import and insightful comments, the model may fail to consider important contextual clues that guide accurate API recommendation, leading to poorer performance. 
Besides, removing the \textit{Bridge Memory} and aggregating the information of all snippets instead results in a relatively slight performance decrease in {\tool}. For example, on API recommendation, the average SR@1 and MRR decrease by 8.12\% and 7.90\% on five baseline models, respectively. The performance drop
can be attributed to the absence of maintaining the continuity of context, which hinders the model's ability to connect related but non-adjacent code elements. 
The results highlight the importance of \textit{Bridge Memory} and \textit{Hint Memory} mechanisms in processing long-range code.

\begin{tcolorbox}[breakable,width=\linewidth-2pt,boxrule=0pt,top=3pt, bottom=3pt, left=4pt,right=4pt, colback=gray!15,colframe=gray!15]
\textbf{Answer to RQ2:} 
Both mechanisms are essential for the performance of {\tool}. Without \textit{Bridge Memory} and \textit{Hint Memory}, the overall performance of {\tool} is decreased by 8.01\% and 13.58\% on API recommendation, respectively.
\end{tcolorbox}

\subsection{Performance Comparison with LLMs (RQ3)}
\begin{table*}[htbp]
\renewcommand\arraystretch{0.9}
\centering
\caption{Performance comparison between {\tool} and large language models.}
\label{tab:comparison_llm}

\begin{tabular}{c|ccccc|ccccc|ccc|ccc}
    \toprule
    \multirow{3}{*}{\textbf{Model}} & \multicolumn{10}{c|}{\textbf{API Recommendation}}  & \multicolumn{6}{c}{\textbf{Vulnerability Detection}}     \\
    \cmidrule{3-10} 
    \cmidrule{13-16} 
       & & \multicolumn{4}{c|}{\textbf{Java}} & \multicolumn{4}{c}{\textbf{Python}} & & &\multicolumn{2}{c|}{\textbf{Java}} & \multicolumn{2}{c}{\textbf{Python}}& \\
            &  & SR@1  & SR@3  & SR@5 & MRR  & SR@1  & SR@3  & SR@5 & MRR & & & Acc. & F1 & Acc.   & F1 &  \\
    \midrule
    CodeGen (7B) & & 36.30    & 38.10  &  41.10  &   0.377   &  36.80 &   37.70  & 33.30  &  0.346  & & & 17.86  & 4.17 &  24.07   &  12.77  &  \\
    ChatGLM (6B) & &  \textbf{63.90}  & 65.50 & 66.50  & 0.647    &  58.80   &   60.90   & 63.00  &   0.601 & & & 44.64   & 11.43   & 38.89  &  10.81 &  \\
    ChatGPT (175B) & & 51.20    & 61.00  & 62.70  &  0.560      & 30.90  & 40.70    &  43.30   & 0.358 & &   & 58.92    & 51.06   & 51.85  &53.57  & \\
    GPT3.5 (175B) & & 56.50   & 56.80  & 57.10    & 0.567     & 56.90  &57.20     &57.20   &0.570  &&  &51.78     &58.46  & 50.50  & 50.90 & \\
    \midrule
    {\tool} (UniXcoder) && 53.55   & \textbf{70.80} &  \textbf{81.38} & \textbf{0.671}   & \textbf{62.04}  & \textbf{75.18 } & \textbf{83.02}   &  \textbf{0.670}  &&  & \textbf{60.71}   & \textbf{66.67}  & \textbf{62.96}  &\textbf{68.75}     &   \\
    \bottomrule
    \end{tabular}
\end{table*}

\noindent \textbf{Experimental Design.} To answer this research question, we utilize four LLMs: CodeGen (CodeGen2.5-7B-instruct)~\cite{Nijkamp2023codegen2}, ChatGLM (ChatGLM3-6B-Base)~\cite{du2022glm}, ChatGPT (gpt-3.5-turbo)~\cite{chatgpt2022} and GPT-3.5 (text-davinci-003)~\cite{gpt3}. For CodeGen and ChatGLM, we download them from HuggingFace Hub~\cite{HuggingFace} and deploy them locally. For ChatGPT and GPT-3.5, we use the public APIs provided by OpenAI. Considering the token limit, we do not provide any examples and only use the task instruction~\cite{DBLP:journals/corr/abs-2311-16169} as the input prompt for LLMs. Due to limited computation resources, we perform each evaluation on the API recommendation task by randomly sampling
1000 instances
from the full test set. We repeat the sampling
process 
three times to mitigate sampling bias,
and report the average results. For vulnerability detection, we employ the full test sets for evaluation.

\noindent \textbf{Results.} Table~\ref{tab:comparison_llm} presents the performance comparison between UniXcoder fine-tuned by {\tool} and various LLMs. The results show that UniXcoder, fine-tuned with {\tool}, outperforms four LLMs in almost all metrics (11/12) on API recommendation and vulnerability detection.

\noindent \textbf{Analyses.} For API recommendation, {\tool} achieves an average SR@5 of 82.20 and an average MRR of 0.670, making an improvement of 27.07\% and 7.60\% compared to the best baseline ChatGLM, respectively. 
As for vulnerability detection, {\tool} 's average Accuracy and F1 score are 61.84 and 67.71, outperforming the best baseline ChatGPT by 11.65\% and 29.42\%, respectively.

The preliminary results indicate that smaller PLMs fine-tuned through {\tool} on task-specific datasets, can outperform larger models with billions of parameters in processing long-range code sequences. This observation is also in agreement with recent work ~\cite{DBLP:conf/acl/GaoZLZW23, DBLP:conf/acl/HsiehLYNFRKLP23}. The possible explanations are: 1) Fine-tuning enables smaller models to concentrate on the tasks they have been trained for, allowing them to perform exceptionally well in specialized domains. This focused training can lead smaller models to exceed the capabilities of larger, more generalized models in these targeted tasks. 2) The datasets chosen for {\tool} fine-tuning, APIBench-C and CrossVul, are likely to be particularly advantageous for the tasks they are designed for. This advantage can potentially boost the performance of smaller models, especially if the larger models have not been fine-tuned on similarly relevant high-quality datasets.



\begin{tcolorbox}[breakable,width=\linewidth-2pt,boxrule=0pt,top=3pt, bottom=3pt, left=4pt,right=4pt, colback=gray!15,colframe=gray!15]
\textbf{Answer to RQ3:} 
In the long-range code scenario, PLMs
fine-tuned with {\tool} on task-specific datasets, demonstrate comparable performance to LLMs.

\end{tcolorbox}

\subsection{Parameter Analysis (RQ4)} \label{sec:para}
\definecolor{c1}{RGB}{255,196,115}
\definecolor{c2}{RGB}{178,37,42} 
\definecolor{c3}{RGB}{103,150,118} 




\pgfplotstableread[row sep=\\,col sep=&]{
	k & java & python   \\
	1 & 53.75 & 53.70   \\
	2 & 57.14 & 55.56  \\
	3 & 55.56 & 55.36 \\
	  4 & 55.36 & 53.70  \\
        5 & 53.57 & 51.35 \\
}\VACKT


\pgfplotstableread[row sep=\\,col sep=&]{
	k & java & python   \\
	1 & 53.57 & 57.40   \\
	2 & 62.50 & 58.62  \\
	3 & 60.71 & 62.96 \\
	  4 & 53.57 & 61.11  \\
        5 & 51.35 & 55.56 \\
}\VACKU


\pgfplotstableread[row sep=\\,col sep=&]{
	k & java & python   \\
	1 & 55.36 & 55.56   \\
	2 & 53.70 & 53.70  \\
	3 & 58.92 & 57.40 \\
	  4 & 51.78 & 57.40  \\
        5 & 53.57 & 50.00 \\
}\VACMT

\pgfplotstableread[row sep=\\,col sep=&]{
	k & java & python  \\
	1 & 66.67 & 68.42   \\
	2 & 68.35 & 69.87 \\
	3 & 71.60 & 70.12 \\
	  4 & 68.23 & 63.49  \\
        5 & 69.76 & 54.23 \\
}\API

\pgfplotstableread[row sep=\\,col sep=&]{
	k & java & python   \\
	1 & 60.71 & 62.96   \\
	2 & 58.92 & 61.53  \\
	3 & 60.71 & 62.96 \\
	  4 & 55.35 & 55.56  \\
        5 & 57.14& 57.40 \\
}\VACMU


\begin{figure*}
    \subfigure[Bridge token number of CodeT5]{
        \begin{tikzpicture}[scale=0.5]
            \huge
            \begin{axis}[
                legend style = {
                    legend columns=2,
                    draw=none,
                },
                xtick = {1,2,3,4,5},
                xticklabels = {1,4,8,16,32},
                ymin=49,ymax=61,
                ytick = {50,52,54,56,58,60},
                mark size=3.5pt, 
                ylabel={\bf Acc.},
                xlabel={\bf $m$},
                every axis plot/.append style={line width = 2.5pt},
                every axis/.append style={line width = 1.6pt},
                ]
                \addplot [mark=diamond,color=c2] table[x=k,y=java]{\VACMT};
                \addplot [mark=pentagon,color=c3] table[x=k,y=python]{\VACMT};
                \legend{Java, Python}
            \end{axis}
        \end{tikzpicture}
        \label{fig:para_T5_m}
	}
    \subfigure[Bridge token number of UniXcoder]{
        \begin{tikzpicture}[scale=0.5]
            \huge
            \begin{axis}[
                legend style = {
                    legend columns=2,
                    draw=none,
                },
                xtick = {1,2,3,4,5},
                xticklabels = {1,4,8,16,32},
                ymin=54,ymax=65,
                ytick = {56,58,60,62,64},
                mark size=3.5pt, 
                ylabel={\bf Acc.},
                xlabel={\bf $m$}, 
                every axis plot/.append style={line width = 2.5pt},
                every axis/.append style={line width = 1.6pt},
                ]
                \addplot [mark=diamond,color=c2] table[x=k,y=java]{\VACMU};
                \addplot [mark=pentagon,color=c3] table[x=k,y=python]{\VACMU};
                \legend{Java, Python}
            \end{axis}
        \end{tikzpicture}
        \label{fig:para_Un_m}
   }
    \subfigure[Retrieved hint number of CodeT5]{
        \begin{tikzpicture}[scale=0.5]
            \huge
            \begin{axis}[
                legend style = {
                    legend columns=2,
                    draw=none,
                },
                xtick = {1,2,3,4,5},
                xticklabels = {8,16,32,64,128},
                ymin=50,ymax=59,
                ytick = {52,54,56,58},
                mark size=3.5pt, 
                ylabel={\bf Acc.},
                xlabel={\bf $K$}, 
                every axis plot/.append style={line width = 2.5pt},
                every axis/.append style={line width = 1.6pt},
                ]
                \addplot [mark=diamond,color=c2] table[x=k,y=java]{\VACKT};
                \addplot [mark=pentagon,color=c3] table[x=k,y=python]{\VACKT};
                \legend{Java, Python}
            \end{axis}
        \end{tikzpicture}
        \label{fig:para_T5_K}
	}
    \subfigure[Retrieved hint number of UniXcoder]{
        \begin{tikzpicture}[scale=0.5]
            \huge
            \begin{axis}[
                legend style = {
                    legend columns=2,
                    draw=none,
                },
                xtick = {1,2,3,4,5},
                xticklabels = {8,16,32,64,128},
                ymin=50,ymax=66,
                ytick = {52, 54,56,58,60,62,64},
                mark size=3.5pt, 
                ylabel={\bf Acc.},
                xlabel={\bf $K$}, 
                every axis plot/.append style={line width = 2.5pt},
                every axis/.append style={line width = 1.6pt},
                ]
                \addplot [mark=diamond,color=c2] table[x=k,y=java]{\VACKU};
                \addplot [mark=pentagon,color=c3] table[x=k,y=python]{\VACKU};
                \legend{Java, Python}
            \end{axis}
        \end{tikzpicture}
         \label{fig:para_Un_K}
   }
 \caption{Parameter analysis of (a)(b) $\mathbf{m}$  and (c)(d) $\mathbf{K}$ with {\tool} (CodeT5) and {\tool} (UniXcoder) for vulnerability detection.
 } 
 \label{fig:para}
\end{figure*}
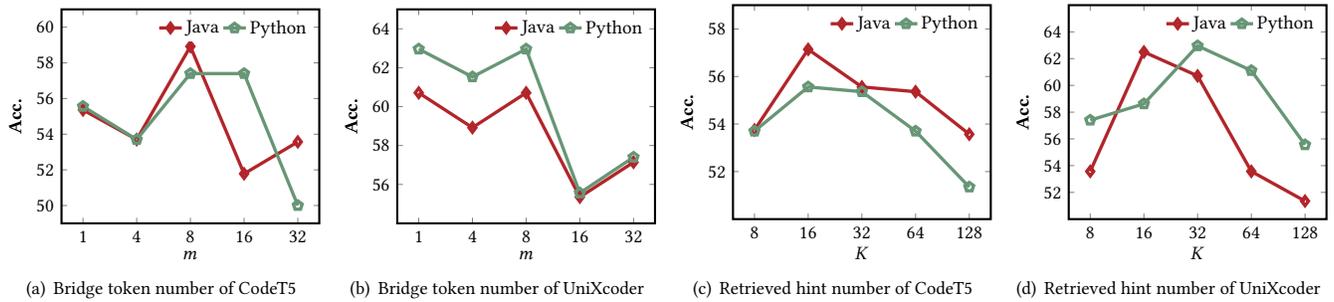

\noindent \textbf{Experimental Design.} To answer this question, we study the impact of two parameters on results, including the bridge token number $m$ which represents the capacity of the model to maintain contextual information across code snippets and retrieved hint number $K$ which indicates how many code
hints are considered for enhancing the current context. We use two {\tool} fine-tuned models, CodeT5 and UniXcoder, and the vulnerability detection task for investigation. In each study, we only vary the parameter that needs to be analyzed and keep other parameters unchanged. 

\noindent \textbf{Results.} Figure~\ref{fig:para} demonstrates the performance of {\tool}(CodeT5) and {\tool}(UniXcoder) on the vulnerability detection task across different numbers of bridge tokens $m$ and retrieved hints $K$.

\begin{figure*}
    \centering
    \includegraphics[scale=0.42]{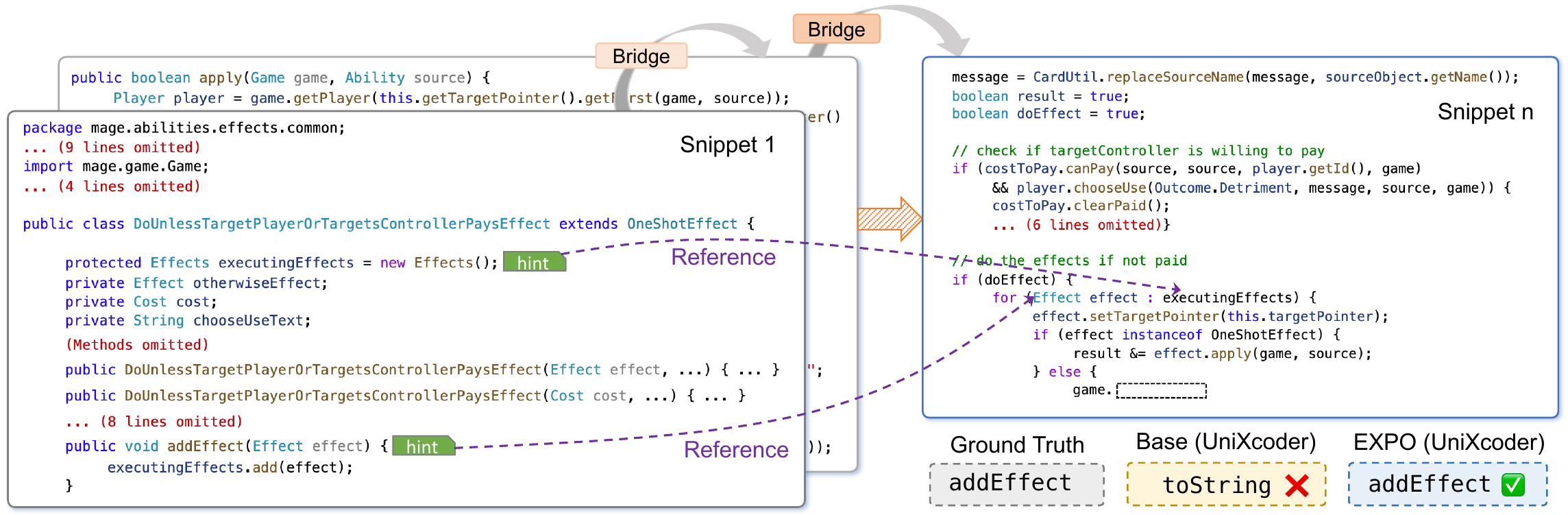}
    \caption{An example of
    predictions of {\tool} (UniXcoder) and the corresponding base model
    for the API recommendation task.
    }
    \label{fig:case}
\end{figure*}

\noindent \textbf{Analyses.} Based on the results, we observe that the performance of {\tool} is largely affected by the number of bridge tokens $m$ and retrieved hints $K$. Specifically, CodeT5 and UniXcoder achieve their best performance when $m$ is fewer
than 8, indicating that a certain number of bridge tokens are necessary for maintaining contextual information between code snippets. However, as $m$ continues to increase, we observe a decline in performance, which surpasses 12.12\% and 12.89\% for Java and Python, respectively. This indicates that too many bridge tokens may introduce noise, resulting in redundancy during the information processing. Regarding retrieved hints, we observe that as $K$ increases, the model's performance initially improves and then diminishes, peaking at $K$ values of 16 or 32. This indicates that a moderate number of hints is necessary when providing global context information. However, too many hints can also harm the performance of {\tool}.
\begin{tcolorbox}[breakable,width=\linewidth-2pt,boxrule=0pt,top=3pt, bottom=3pt, left=4pt,right=4pt, colback=gray!15,colframe=gray!15]
\textbf{Answer to RQ4:}
Selecting an appropriate number of bridge tokens (e.g., 8 for {\tool}) and retrieved hints (16 or 32 for {\tool}) is crucial for the performance of {\tool}.
\end{tcolorbox}

\section{Discussion} \label{sec:discussion}
    \subsection{Why does {\tool} work?}

The effectiveness of {\tool} mainly comes from innovative dual memory mechanisms: \textit{Bridge Memory} and \textit{Hint Memory}, which together enhance the model's capabilities in understanding long-range code sequences. The following analyses of API recommendation and vulnerability detection cases, as demonstrated in Figure~\ref{fig:case} and Section~\ref{sec:motivation}, present how {\tool} achieves its performance.

\subsubsection{Bridge Memory for maintaining contextual continuity.} {\it Bridge Memory} ensures that as the code is parsed into snippets, the context is not lost. It does this by maintaining a ``bridge'' that carries context from one snippet to the next. 
For example, for API recommendation, the case shown in Figure~\ref{fig:case} demonstrates a Java code snippet for which the base model, UniXcoder, incorrectly suggests the {\tt toString} method while the ground truth is the {\tt addEffect} method. This error can be attributed to the base model's limited view of the code context, which focuses on the immediate code snippet without fully understanding the broader context. While for {\tool}, as the bridge tokens carry forward the context from earlier snippets where {\tt addEffect} is defined/used, it provides a continuous understanding that informs the model that {\tt addEffect} is a relevant API call within this context. 
For vulnerability detection, as shown in Figure~\ref{fig:motivation}, functions for safely initializing and updating arrays were defined early in the code, but potential overflow risks appeared in later parts. Traditional PLMs might consider the earlier code snippets as safe, failing to detect vulnerabilities in the {\tt array\_operation} method.
\textit{Bridge Memory} connects the early safe context with later unsafe operations, even if they are separated by hundreds of lines in the code. Through this cross-snippet information transfer, {\tool} can identify deviations from the initially established safety pattern and detect possible buffer overflow vulnerabilities.

\subsubsection{Hint Memory for memorizing key information.} {\it Hint Memory} is responsible for retaining key information throughout the code that is vital for understanding the code's functionality, such as imports, global variable declarations, and comments. 
For API recommendation, as illustrated in Figure~\ref{fig:case}, \textit{Hint memory} further captures and stores attention key-value pairs for tokens like {\tt addEffect} in a centralized hint bank. When the model encounters a situation that requires an API call, the $k$NN attention layer retrieves these pairs, bringing the global context of previous code snippets to the current analysis. This means that despite the {\tt addEffect} method being mentioned earlier in the code, the model can recall its significance and correctly suggest it when needed. 
For vulnerability detection, as shown in Figure~\ref{fig:motivation}, \textit{Hint Memory} tracks the global declaration of {\tt BUFFER\_SIZE} and sets a crucial constraint for all array operations. When {\tool} analyzes the {\tt array\_operation} method, which is far from the initial definition.  \textit{Hint Memory} helps the model recall the predefined maximum array size, aiding in assessing the safety of the operation and identifying potential overflow risks.

By combining the contextual tracking of \textit{Bridge Memory} and the global scope retention of \textit{Hint Memory}, {\tool} mitigates the shortcomings of existing pre-trained language models in handling long-range dependencies within source code. This makes {\tool} particularly adept at tasks like API recommendation and vulnerability detection, where understanding the broader scope and finer details of the code is essential.

\subsection{Impact Analysis of Code Length}
\definecolor{c1}{RGB}{255,196,115} 
\definecolor{c2}{RGB}{178,37,42} 

\pgfplotstableread[row sep=\\,col sep=&]{
    datasets  & java & python   \\
    2 & 50.00  & 50.00     \\ 
    4 & 53.57 & 59.25   \\
    6 & 60.71 & 62.96  \\
    8 & 63.29 & 70.37  \\
}\ACC

\pgfplotstableread[row sep=\\,col sep=&]{
    datasets  & java & python   \\
    2 & 60.16 & 64.94    \\ 
    4 & 63.29 & 66.05   \\
    6 & 66.67 & 68.75  \\
    8 & 71.25 & 75.00  \\
}\FO

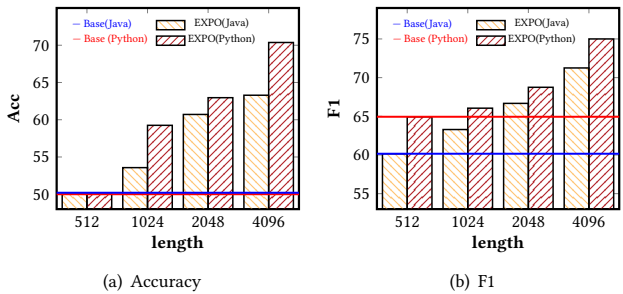
\begin{figure}[tb]	
	\subfigure[Accuracy]{
        \huge
    	\begin{tikzpicture}[scale=0.47]
    		\begin{axis}[
    	    	ybar=0pt,
    			bar width=0.7cm,
                    xmin=1.0,xmax=9.0,
                    xtick=data,	
                    xticklabels={512,1024,2048,4096},
                legend style = {
				    legend columns = 1,
				    draw=none,
                        at={(0.85,0.98)},
				},
				legend image code/.code={
                    \draw [#1] (0cm,-0.18cm) rectangle (0.8cm,0.08cm); },
    			ytick = {50, 55, 60, 65, 70},
    			ymin=48,ymax=75,
    			tick align=inside,
   			  every axis plot/.append style={line width = 1.2pt},
    			every axis/.append style={line width = 1.8pt},
    			ylabel={\textbf{Acc}},
                    xlabel={\textbf{length}}
    			]
    			\addplot[pattern=north west lines, pattern color=c1] table[x=datasets,y=java]{\ACC};
    			\addplot[pattern = north east lines ,pattern color=c2] table[x=datasets,y=python]{\ACC};
    			\legend{\normalsize {\tool}(Java), \normalsize {\tool}(Python)}
                     \draw [red, ultra thick] (axis cs: \pgfkeysvalueof{/pgfplots/xmin},50) -- (axis cs: \pgfkeysvalueof{/pgfplots/xmax},50);
                     \draw [blue, ultra thick] (axis cs: \pgfkeysvalueof{/pgfplots/xmin},50.2) -- (axis cs: \pgfkeysvalueof{/pgfplots/xmax},50.2);
                    \node [below] at (axis cs:2.5,74.5) { \normalsize \textcolor{blue}{--- Base(Java)}};
                    \node [below] at (axis cs:2.77,72.5) { \normalsize \textcolor{red}{--- Base (Python)}};
    		\end{axis}
    	\end{tikzpicture}
    	\label{fig:res-para-shs}
	}
	\subfigure[F1]{
        \huge
    	\begin{tikzpicture}[scale=0.47]
    		\begin{axis}[
    	    	ybar=0pt,
    			bar width=0.7cm,
                    xmin=1.0,xmax=9.0,
                    xtick=data,	
                    xticklabels={512,1024,2048,4096},
                legend style = {
				    legend columns=1,
				    draw=none,
                        at={(0.85,0.98)},
				},
				legend image code/.code={
                    \draw [#1] (0cm,-0.18cm) rectangle (0.8cm,0.08cm); },
    			ytick = {55, 60, 65, 70, 75},
    			ymin=53, ymax=79,
    			tick align=inside,
   			  every axis plot/.append style={line width = 1.2pt},
    			every axis/.append style={line width = 1.8pt},
    			ylabel= { \bf F1},
                    xlabel={\textbf{length}}
    			]
    			\addplot[pattern=north west lines, pattern color=c1] table[x=datasets,y=java]{\FO};
    			\addplot[pattern = north east lines ,pattern color=c2] table[x=datasets,y=python]{\FO};
    	\legend{\normalsize {\tool}(Java), \normalsize {\tool}(Python)}
                    \draw [blue, ultra thick] (axis cs: \pgfkeysvalueof{/pgfplots/xmin},60.16) -- (axis cs: \pgfkeysvalueof{/pgfplots/xmax},60.16);
                    \draw [red, ultra thick] (axis cs: \pgfkeysvalueof{/pgfplots/xmin},64.94) -- (axis cs: \pgfkeysvalueof{/pgfplots/xmax},64.94);
                    \node  [below] at (axis cs:2.5,78.5) { \normalsize \textcolor{blue}{--- Base(Java)}};
                    \node[below] at (axis cs:2.77,76.5) {  \normalsize \textcolor{red}{--- Base (Python)}};
    		\end{axis}
    	\end{tikzpicture}
    	\label{fig:res-para-shs}
	}
        \caption{The performance of {\tool} (UniXcoder) with differing input code lengths for vulnerability detection. The bars and lines indicate the results of {\tool} and base models, respectively.}
        \label{fig:dis_length}
\end{figure} 
In this section, we explore how varying the length of the input code affects the performance of our model and its inference time. We use vulnerability detection for this study, and the results are presented in Figure~\ref{fig:dis_length} and Table~\ref{tab:dis_time}.
We find that, for both Java and Python, {\tool} shows improved Accuracy and F1 score as the length of the code increases and performance peaks at code lengths of 4096 tokens. This suggests \textit{Bridge Memory} and \textit{Hint Memory} mechanisms enable EXPO can not only thread individual code snippets but also incorporate relevant code structures across the entire long source code. This improves its understanding of longer code, leading to consistently superior performance compared to baseline models.
Regarding inference time, there is a clear trend: longer inputs result in longer processing times, which aligns with the expectation that more tokens require more computation effort. Nonetheless, the rise in inference time is relatively modest. For example, for Python, the time increases from 0.018s to 0.071s for tokens ranging from 512 to 4096, which indicates a less than five-fold increase in time despite an eight-fold increase in the input length.
According to Figure~\ref{fig:dis_length} and Table~\ref{tab:dis_time}, we can achieve that {\tool} is effective for long-range code input with acceptable time cost 
on vulnerability detection, which is critical for practical application.



\subsection{Threats to Validity}

We have identified the following major threats to validity:

\subsubsection{Base Models.} 
In this study, we have chosen five widely-used PLMs for evaluation. To comprehensively evaluate the performance of {\tool}, it would be beneficial to include additional PLMs, such as PLBART~\cite{DBLP:conf/naacl/AhmadCRC21}, and also consider non-pre-trained models like Transformer. Additionally, while {\tool} is also compatible with LLMs, this paper does not apply {\tool} to extend LLMs due to constraints of resources and time. In
future work, we plan to explore and validate the effectiveness of our proposed approach on these models.

\subsubsection{Evaluation tasks.}
In this work, we select two popular code intelligence tasks to evaluate {\tool}, including API recommendation and vulnerability detection, and experiment with two widely-used programming languages, i.e., Java and Python. Although {\tool} shows superior performance on these tasks, other important tasks such as code search~\cite{DBLP:conf/sigsoft/CambroneroLKS019,DBLP:journals/nn/GuLGWZXL21} and code summarization~\cite{DBLP:conf/iwpc/HuLXLJ18, DBLP:conf/acl/AhmadCRC20} are not evaluated in our experiment. In the future, we will validate {\tool} on more code intelligence tasks and more programming languages.

\subsubsection{Prompt design for LLMs.} In this work, we solely rely on task instructions as input prompts for LLMs without offering examples, constrained by token limits. This method might not harness the full potential of LLMs. In the future, we will explore LLMs' capabilities on the two tasks with diverse prompts. 

\section{Related Work}  \label{sec:related}
\begin{table}[tb]
\centering
\renewcommand\arraystretch{1.2}
\caption{The inference time cost of {\tool} (UniXcoder) with varying code lengths for vulnerability detection.}
\label{tab:dis_time}
\begin{tabular}{c|cccc}
\toprule 
\multirow{2}{*}{\textbf{Code Length}}  &  \multicolumn{4}{c}{\textbf{Time Cost (Per Instance)}} \\ 
{} & 512  & 1024  & 2048 & 4096 \\
\midrule
Java   & 0.018s  & 0.037s & 0.053s & 0.082s \\ 
Python & 0.018s  & 0.030s & 0.042s & 0.071s \\ 
\bottomrule 
\end{tabular}
\end{table}
 
\subsection{Pre-trained Language Models}
Pre-trained language models can achieve general-purpose language comprehension and generation by unsupervised learning on large-scale unlabeled text corpora. For example, BERT~\cite{DBLP:conf/naacl/DevlinCLT19} is trained to acquire the contextual knowledge through masked language model objectives, and subsequently fine-tuned for various down-streaming tasks. 
T5~\cite{T5} utilizes a text-to-text framework for pre-training, acquiring the ability to transform input text into target text across various NLP tasks, thereby enabling versatile and cohesive natural language processing capabilities.
In contrast, GPT utilizes autoregressive language modeling to predict successive words in a sequence~\cite{gpt1}, enabling adaptation to diverse tasks by modifying the input format~\cite{gpt2,gpt3}.
Over the past year, the significant success of ChatGPT has brought widespread attention to LLMs~\cite{DBLP:journals/corr/abs-2311-16989}. The powerful capabilities of LLMs allow them to achieve excellent performance without requiring fine-tuning for downstream tasks~\cite{gpt4,llama2}.

\subsection{Code Representation Learning}
Code representation learning aims to encode code fragments into vector representations that can be used in various downstreaming tasks, such as API recommendation~\cite{DBLP:conf/icse/WeiHH0022, DBLP:journals/tse/ZhouYCHMG22} and vulnerability detection~\cite{DBLP:conf/nips/ZhouLSD019,DBLP:journals/tse/ChakrabortyKDR22,DBLP:conf/sigsoft/Li0N21}.
For example, ASTNN~\cite{astnn} uses Recurrent Neural Networks (RNNs) to encode Abstract Syntax Trees (ASTs) for learning the code representations.
Inspired by the success of PLMs
in the field of NLP, the models pre-trained on a large amount of code
are employed for code representation learning~\cite{DBLP:conf/iclr/GuoRLFT0ZDSFTDC21}. CodeBERT~\cite{DBLP:conf/emnlp/FengGTDFGS0LJZ20} is pre-trained on unlabeled
six programming languages through masked language modeling objectives.
CodeT5~\cite{DBLP:conf/emnlp/0034WJH21} adopts identifier-aware denoising pre-training to fuse more code-specific structural information into the model. Also, UniXcoder~\cite{DBLP:conf/acl/GuoLDW0022} leverages
ASTs and code comments to enhance code representations.
Subsequently, other code language models
are proposed
\cite{DBLP:journals/corr/abs-2303-17568,DBLP:journals/corr/abs-2305-02309}.

Long code segments are quite common in the field of software engineering, presenting a challenge in capturing the intricate context and dependencies. General pre-trained code models struggle with handling long code due to the complex context and dependencies~\cite{DBLP:conf/emnlp/ClementLLTDDSS21}.
LongCoder~\cite{DBLP:conf/icml/GuoXD0M23} is a long-range PLM
with sparse attention mechanism specifically tailored for the code completion task. Different from LongCoder, {\tool} is the first general framework to enhance the existing pre-trained code model for long-range code representation learning. 
Furthermore, {\tool} fuses more code-specific information to adapt various code-related tasks.

\subsection{API Recommendation}
The works usually utilize traditional statistical methods to capture API usage patterns from API co-occurrence or leverage deep learning models to automatically learn the potential usage patterns from a large code corpus. 
Zhong \etal propose MAPO~\cite{DBLP:conf/ecoop/ZhongXZPM09} to cluster and mine API usage patterns from open source repositories, then recommend the relevant usage patterns to developers. 
Fowkes \etal propose PAM~\cite{DBLP:conf/sigsoft/FowkesS16} to mine API usage patterns through an almost parameter-free probabilistic algorithm and uses them to recommend APIs. 
Nguyen \etal propose a graph-based language model GraLan~\cite{DBLP:conf/icse/NguyenN15} to recommend API usages. Besides, some work leverages API recommendation technologies for various tasks. For example, Wei \etal~\cite{DBLP:conf/sigsoft/0003X023} aim to enhance the efficiency of Automated Program Repair (APR) task by proposing several memorization techniques to reduce the frequency of invoking the Completion Engine. 

\subsection{Vulnerability Detection}
The techniques can be classified
into two categories: sequence-based and graph-based methods. For sequence-based methods, Russell \etal~\cite{DBLP:conf/icmla/RussellKHLHOEM18} utilize Convolutional Neural Networks and Recurrent Neural Networks to fuse different features from function-level source code. SySeVR~\cite{DBLP:journals/tdsc/0027ZX0ZC22} extracts code gadgets by traversing AST generated from code and also uses a Bi-LSTM network. For graph-based methods, Devign~\cite{DBLP:conf/nips/ZhouLSD019} adds natural code sequence to the code property graph and leverages the Gated Graph Neutral
Networks, which preserve the programming logic
of the source code. AMPLE~\cite{DBLP:conf/icse/WenCGZZL23} shrinks the code structure graphs to reduce the distances between nodes and designs an enhanced graph representation learning. PILOT~\cite{DBLP:conf/kbse/WenWGWLG23} consists of a distance-aware label selection for generating pseudo labels and a mixed-supervision representation learning module to alleviate the influence of noise. 

\section{Conclusion} \label{sec:conclusion}
    In this paper, we investigate pre-trained language models for code intelligence tasks in the long-range code scenario. To mitigate the challenges of contextual continuity maintenance and key information memorization, we propose a general framework {\tool}, empowering pre-trained language models for effective long-range code modeling. The core of {\tool} is a dual-memory mechanism, including \textit{Bridge Memory} for recurrently transferring information across code snippets, and \textit{Hint Memory} for storing and retrieving global code elements. The evaluation on
two common tasks demonstrate the effectiveness of {\tool} to effectively model long-range code. In the future, we will apply {\tool} to extend more pre-trained language models and validate them on more code intelligence tasks. 


\section{Data Availability} 
We release our source code and data at \url{https://anonymous.4open.science/r/EXPO/}.

\begin{acks}
We thank all the anonymous reviewers. This research is supported by Natural Science Foundation of Guangdong Province (Project No. 2023A1515011959), Shenzhen-Hong Kong Jointly Funded Project (Category A, No. SGDX20230116091246007), Shenzhen Basic Research (General Project No. JCYJ20220531095214031), Shenzhen International Science and Technology Cooperation Project (No. GJHZ20220913143008015), and the Major Key Project of PCL (Grant No. PCL2022A03).
\end{acks}

\normalem
\bibliographystyle{ACM-Reference-Format}
\balance
\bibliography{ref}


\begin{thebibliography}{67}


\ifx \showCODEN    \undefined \def \showCODEN     #1{\unskip}     \fi
\ifx \showDOI      \undefined \def \showDOI       #1{#1}\fi
\ifx \showISBNx    \undefined \def \showISBNx     #1{\unskip}     \fi
\ifx \showISBNxiii \undefined \def \showISBNxiii  #1{\unskip}     \fi
\ifx \showISSN     \undefined \def \showISSN      #1{\unskip}     \fi
\ifx \showLCCN     \undefined \def \showLCCN      #1{\unskip}     \fi
\ifx \shownote     \undefined \def \shownote      #1{#1}          \fi
\ifx \showarticletitle \undefined \def \showarticletitle #1{#1}   \fi
\ifx \showURL      \undefined \def \showURL       {\relax}        \fi
\providecommand\bibfield[2]{#2}
\providecommand\bibinfo[2]{#2}
\providecommand\natexlab[1]{#1}
\providecommand\showeprint[2][]{arXiv:#2}

\bibitem[Ahmad et~al\mbox{.}(2020)]%
        {DBLP:conf/acl/AhmadCRC20}
\bibfield{author}{\bibinfo{person}{Wasi~Uddin Ahmad}, \bibinfo{person}{Saikat Chakraborty}, \bibinfo{person}{Baishakhi Ray}, {and} \bibinfo{person}{Kai{-}Wei Chang}.} \bibinfo{year}{2020}\natexlab{}.
\newblock \showarticletitle{A Transformer-based Approach for Source Code Summarization}. In \bibinfo{booktitle}{\emph{Proceedings of the 58th Annual Meeting of the Association for Computational Linguistics, {ACL} 2020, Online, July 5-10, 2020}}, \bibfield{editor}{\bibinfo{person}{Dan Jurafsky}, \bibinfo{person}{Joyce Chai}, \bibinfo{person}{Natalie Schluter}, {and} \bibinfo{person}{Joel~R. Tetreault}} (Eds.). \bibinfo{publisher}{Association for Computational Linguistics}, \bibinfo{pages}{4998--5007}.
\newblock


\bibitem[Ahmad et~al\mbox{.}(2021)]%
        {DBLP:conf/naacl/AhmadCRC21}
\bibfield{author}{\bibinfo{person}{Wasi~Uddin Ahmad}, \bibinfo{person}{Saikat Chakraborty}, \bibinfo{person}{Baishakhi Ray}, {and} \bibinfo{person}{Kai{-}Wei Chang}.} \bibinfo{year}{2021}\natexlab{}.
\newblock \showarticletitle{Unified Pre-training for Program Understanding and Generation}. In \bibinfo{booktitle}{\emph{Proceedings of the 2021 Conference of the North American Chapter of the Association for Computational Linguistics: Human Language Technologies, {NAACL-HLT} 2021, Online, June 6-11, 2021}}, \bibfield{editor}{\bibinfo{person}{Kristina Toutanova}, \bibinfo{person}{Anna Rumshisky}, \bibinfo{person}{Luke Zettlemoyer}, \bibinfo{person}{Dilek Hakkani{-}T{\"{u}}r}, \bibinfo{person}{Iz~Beltagy}, \bibinfo{person}{Steven Bethard}, \bibinfo{person}{Ryan Cotterell}, \bibinfo{person}{Tanmoy Chakraborty}, {and} \bibinfo{person}{Yichao Zhou}} (Eds.). \bibinfo{publisher}{Association for Computational Linguistics}, \bibinfo{pages}{2655--2668}.
\newblock


\bibitem[Bacchelli and Bird(2013)]%
        {DBLP:conf/icse/BacchelliB13}
\bibfield{author}{\bibinfo{person}{Alberto Bacchelli} {and} \bibinfo{person}{Christian Bird}.} \bibinfo{year}{2013}\natexlab{}.
\newblock \showarticletitle{Expectations, outcomes, and challenges of modern code review}. In \bibinfo{booktitle}{\emph{35th International Conference on Software Engineering, {ICSE} '13, San Francisco, CA, USA, May 18-26, 2013}}, \bibfield{editor}{\bibinfo{person}{David Notkin}, \bibinfo{person}{Betty H.~C. Cheng}, {and} \bibinfo{person}{Klaus Pohl}} (Eds.). \bibinfo{publisher}{{IEEE} Computer Society}, \bibinfo{pages}{712--721}.
\newblock


\bibitem[Brown et~al\mbox{.}(2020)]%
        {gpt3}
\bibfield{author}{\bibinfo{person}{Tom Brown}, \bibinfo{person}{Benjamin Mann}, \bibinfo{person}{Nick Ryder}, \bibinfo{person}{Melanie Subbiah}, \bibinfo{person}{Jared~D Kaplan}, \bibinfo{person}{Prafulla Dhariwal}, \bibinfo{person}{Arvind Neelakantan}, \bibinfo{person}{Pranav Shyam}, \bibinfo{person}{Girish Sastry}, \bibinfo{person}{Amanda Askell}, {et~al\mbox{.}}} \bibinfo{year}{2020}\natexlab{}.
\newblock \showarticletitle{Language models are few-shot learners}.
\newblock \bibinfo{journal}{\emph{Advances in neural information processing systems}}  \bibinfo{volume}{33} (\bibinfo{year}{2020}), \bibinfo{pages}{1877--1901}.
\newblock


\bibitem[Butler(2012)]%
        {DBLP:conf/icse/Butler12}
\bibfield{author}{\bibinfo{person}{Simon Butler}.} \bibinfo{year}{2012}\natexlab{}.
\newblock \showarticletitle{Mining Java class identifier naming conventions}. In \bibinfo{booktitle}{\emph{34th International Conference on Software Engineering, {ICSE} 2012, June 2-9, 2012, Zurich, Switzerland}}, \bibfield{editor}{\bibinfo{person}{Martin Glinz}, \bibinfo{person}{Gail~C. Murphy}, {and} \bibinfo{person}{Mauro Pezz{\`{e}}}} (Eds.). \bibinfo{publisher}{{IEEE} Computer Society}, \bibinfo{pages}{1641--1643}.
\newblock


\bibitem[Cambronero et~al\mbox{.}(2019)]%
        {DBLP:conf/sigsoft/CambroneroLKS019}
\bibfield{author}{\bibinfo{person}{Jos{\'{e}} Cambronero}, \bibinfo{person}{Hongyu Li}, \bibinfo{person}{Seohyun Kim}, \bibinfo{person}{Koushik Sen}, {and} \bibinfo{person}{Satish Chandra}.} \bibinfo{year}{2019}\natexlab{}.
\newblock \showarticletitle{When deep learning met code search}. In \bibinfo{booktitle}{\emph{Proceedings of the {ACM} Joint Meeting on European Software Engineering Conference and Symposium on the Foundations of Software Engineering, {ESEC/SIGSOFT} {FSE} 2019, Tallinn, Estonia, August 26-30, 2019}}, \bibfield{editor}{\bibinfo{person}{Marlon Dumas}, \bibinfo{person}{Dietmar Pfahl}, \bibinfo{person}{Sven Apel}, {and} \bibinfo{person}{Alessandra Russo}} (Eds.). \bibinfo{publisher}{{ACM}}, \bibinfo{pages}{964--974}.
\newblock


\bibitem[Casalnuovo et~al\mbox{.}(2020)]%
        {DBLP:conf/icse/CasalnuovoBDDM20}
\bibfield{author}{\bibinfo{person}{Casey Casalnuovo}, \bibinfo{person}{Earl~T. Barr}, \bibinfo{person}{Santanu~Kumar Dash}, \bibinfo{person}{Prem Devanbu}, {and} \bibinfo{person}{Emily Morgan}.} \bibinfo{year}{2020}\natexlab{}.
\newblock \showarticletitle{A theory of dual channel constraints}. In \bibinfo{booktitle}{\emph{{ICSE-NIER} 2020: 42nd International Conference on Software Engineering, New Ideas and Emerging Results, Seoul, South Korea, 27 June - 19 July, 2020}}, \bibfield{editor}{\bibinfo{person}{Gregg Rothermel} {and} \bibinfo{person}{Doo{-}Hwan Bae}} (Eds.). \bibinfo{publisher}{{ACM}}, \bibinfo{pages}{25--28}.
\newblock


\bibitem[Chakraborty et~al\mbox{.}(2022)]%
        {DBLP:journals/tse/ChakrabortyKDR22}
\bibfield{author}{\bibinfo{person}{Saikat Chakraborty}, \bibinfo{person}{Rahul Krishna}, \bibinfo{person}{Yangruibo Ding}, {and} \bibinfo{person}{Baishakhi Ray}.} \bibinfo{year}{2022}\natexlab{}.
\newblock \showarticletitle{Deep Learning Based Vulnerability Detection: Are We There Yet?}
\newblock \bibinfo{journal}{\emph{{IEEE} Trans. Software Eng.}} \bibinfo{volume}{48}, \bibinfo{number}{9} (\bibinfo{year}{2022}), \bibinfo{pages}{3280--3296}.
\newblock


\bibitem[ChatGPT(2022)]%
        {chatgpt2022}
\bibfield{author}{\bibinfo{person}{ChatGPT}.} \bibinfo{year}{2022}\natexlab{}.
\newblock
\newblock
\urldef\tempurl%
\url{https://chat.openai.com/}
\showURL{%
\tempurl}


\bibitem[Chen et~al\mbox{.}(2023b)]%
        {DBLP:journals/corr/abs-2311-16989}
\bibfield{author}{\bibinfo{person}{Hailin Chen}, \bibinfo{person}{Fangkai Jiao}, \bibinfo{person}{Xingxuan Li}, \bibinfo{person}{Chengwei Qin}, \bibinfo{person}{Mathieu Ravaut}, \bibinfo{person}{Ruochen Zhao}, \bibinfo{person}{Caiming Xiong}, {and} \bibinfo{person}{Shafiq Joty}.} \bibinfo{year}{2023}\natexlab{b}.
\newblock \showarticletitle{ChatGPT's One-year Anniversary: Are Open-Source Large Language Models Catching up?}
\newblock \bibinfo{journal}{\emph{CoRR}}  \bibinfo{volume}{abs/2311.16989} (\bibinfo{year}{2023}).
\newblock


\bibitem[Chen et~al\mbox{.}(2023a)]%
        {DBLP:journals/tse/ChenGRP0L23}
\bibfield{author}{\bibinfo{person}{Yujia Chen}, \bibinfo{person}{Cuiyun Gao}, \bibinfo{person}{Xiaoxue Ren}, \bibinfo{person}{Yun Peng}, \bibinfo{person}{Xin Xia}, {and} \bibinfo{person}{Michael~R. Lyu}.} \bibinfo{year}{2023}\natexlab{a}.
\newblock \showarticletitle{{API} Usage Recommendation Via Multi-View Heterogeneous Graph Representation Learning}.
\newblock \bibinfo{journal}{\emph{{IEEE} Trans. Software Eng.}} \bibinfo{volume}{49}, \bibinfo{number}{5} (\bibinfo{year}{2023}), \bibinfo{pages}{3289--3304}.
\newblock


\bibitem[Clement et~al\mbox{.}(2021)]%
        {DBLP:conf/emnlp/ClementLLTDDSS21}
\bibfield{author}{\bibinfo{person}{Colin~B. Clement}, \bibinfo{person}{Shuai Lu}, \bibinfo{person}{Xiaoyu Liu}, \bibinfo{person}{Michele Tufano}, \bibinfo{person}{Dawn Drain}, \bibinfo{person}{Nan Duan}, \bibinfo{person}{Neel Sundaresan}, {and} \bibinfo{person}{Alexey Svyatkovskiy}.} \bibinfo{year}{2021}\natexlab{}.
\newblock \showarticletitle{Long-Range Modeling of Source Code Files with eWASH: Extended Window Access by Syntax Hierarchy}. In \bibinfo{booktitle}{\emph{Proceedings of the 2021 Conference on Empirical Methods in Natural Language Processing, {EMNLP} 2021, Virtual Event / Punta Cana, Dominican Republic, 7-11 November, 2021}}, \bibfield{editor}{\bibinfo{person}{Marie{-}Francine Moens}, \bibinfo{person}{Xuanjing Huang}, \bibinfo{person}{Lucia Specia}, {and} \bibinfo{person}{Scott~Wen{-}tau Yih}} (Eds.). \bibinfo{publisher}{Association for Computational Linguistics}, \bibinfo{pages}{4713--4722}.
\newblock


\bibitem[Devlin et~al\mbox{.}(2019)]%
        {DBLP:conf/naacl/DevlinCLT19}
\bibfield{author}{\bibinfo{person}{Jacob Devlin}, \bibinfo{person}{Ming{-}Wei Chang}, \bibinfo{person}{Kenton Lee}, {and} \bibinfo{person}{Kristina Toutanova}.} \bibinfo{year}{2019}\natexlab{}.
\newblock \showarticletitle{{BERT:} Pre-training of Deep Bidirectional Transformers for Language Understanding}. In \bibinfo{booktitle}{\emph{Proceedings of the 2019 Conference of the North American Chapter of the Association for Computational Linguistics: Human Language Technologies, {NAACL-HLT} 2019, Minneapolis, MN, USA, June 2-7, 2019, Volume 1 (Long and Short Papers)}}, \bibfield{editor}{\bibinfo{person}{Jill Burstein}, \bibinfo{person}{Christy Doran}, {and} \bibinfo{person}{Thamar Solorio}} (Eds.). \bibinfo{publisher}{Association for Computational Linguistics}, \bibinfo{pages}{4171--4186}.
\newblock


\bibitem[Du et~al\mbox{.}(2022)]%
        {du2022glm}
\bibfield{author}{\bibinfo{person}{Zhengxiao Du}, \bibinfo{person}{Yujie Qian}, \bibinfo{person}{Xiao Liu}, \bibinfo{person}{Ming Ding}, \bibinfo{person}{Jiezhong Qiu}, \bibinfo{person}{Zhilin Yang}, {and} \bibinfo{person}{Jie Tang}.} \bibinfo{year}{2022}\natexlab{}.
\newblock \showarticletitle{GLM: General Language Model Pretraining with Autoregressive Blank Infilling}. In \bibinfo{booktitle}{\emph{Proceedings of the 60th Annual Meeting of the Association for Computational Linguistics (Volume 1: Long Papers)}}. \bibinfo{pages}{320--335}.
\newblock


\bibitem[Feng et~al\mbox{.}(2020)]%
        {DBLP:conf/emnlp/FengGTDFGS0LJZ20}
\bibfield{author}{\bibinfo{person}{Zhangyin Feng}, \bibinfo{person}{Daya Guo}, \bibinfo{person}{Duyu Tang}, \bibinfo{person}{Nan Duan}, \bibinfo{person}{Xiaocheng Feng}, \bibinfo{person}{Ming Gong}, \bibinfo{person}{Linjun Shou}, \bibinfo{person}{Bing Qin}, \bibinfo{person}{Ting Liu}, \bibinfo{person}{Daxin Jiang}, {and} \bibinfo{person}{Ming Zhou}.} \bibinfo{year}{2020}\natexlab{}.
\newblock \showarticletitle{CodeBERT: {A} Pre-Trained Model for Programming and Natural Languages}. In \bibinfo{booktitle}{\emph{Findings of the Association for Computational Linguistics: {EMNLP} 2020, Online Event, 16-20 November 2020}} \emph{(\bibinfo{series}{Findings of {ACL}}, Vol.~\bibinfo{volume}{{EMNLP} 2020})}, \bibfield{editor}{\bibinfo{person}{Trevor Cohn}, \bibinfo{person}{Yulan He}, {and} \bibinfo{person}{Yang Liu}} (Eds.). \bibinfo{publisher}{Association for Computational Linguistics}, \bibinfo{pages}{1536--1547}.
\newblock


\bibitem[Fowkes and Sutton(2016)]%
        {DBLP:conf/sigsoft/FowkesS16}
\bibfield{author}{\bibinfo{person}{Jaroslav~M. Fowkes} {and} \bibinfo{person}{Charles Sutton}.} \bibinfo{year}{2016}\natexlab{}.
\newblock \showarticletitle{Parameter-free probabilistic {API} mining across GitHub}. In \bibinfo{booktitle}{\emph{Proceedings of the 24th {ACM} {SIGSOFT} International Symposium on Foundations of Software Engineering, {FSE} 2016, Seattle, WA, USA, November 13-18, 2016}}, \bibfield{editor}{\bibinfo{person}{Thomas Zimmermann}, \bibinfo{person}{Jane Cleland{-}Huang}, {and} \bibinfo{person}{Zhendong Su}} (Eds.). \bibinfo{publisher}{{ACM}}, \bibinfo{pages}{254--265}.
\newblock


\bibitem[Gao et~al\mbox{.}(2023)]%
        {DBLP:conf/acl/GaoZLZW23}
\bibfield{author}{\bibinfo{person}{Ze{-}Feng Gao}, \bibinfo{person}{Kun Zhou}, \bibinfo{person}{Peiyu Liu}, \bibinfo{person}{Wayne~Xin Zhao}, {and} \bibinfo{person}{Ji{-}Rong Wen}.} \bibinfo{year}{2023}\natexlab{}.
\newblock \showarticletitle{Small Pre-trained Language Models Can be Fine-tuned as Large Models via Over-Parameterization}. In \bibinfo{booktitle}{\emph{Proceedings of the 61st Annual Meeting of the Association for Computational Linguistics (Volume 1: Long Papers), {ACL} 2023, Toronto, Canada, July 9-14, 2023}}, \bibfield{editor}{\bibinfo{person}{Anna Rogers}, \bibinfo{person}{Jordan~L. Boyd{-}Graber}, {and} \bibinfo{person}{Naoaki Okazaki}} (Eds.). \bibinfo{publisher}{Association for Computational Linguistics}, \bibinfo{pages}{3819--3834}.
\newblock


\bibitem[Gu et~al\mbox{.}(2021)]%
        {DBLP:journals/nn/GuLGWZXL21}
\bibfield{author}{\bibinfo{person}{Wenchao Gu}, \bibinfo{person}{Zongjie Li}, \bibinfo{person}{Cuiyun Gao}, \bibinfo{person}{Chaozheng Wang}, \bibinfo{person}{Hongyu Zhang}, \bibinfo{person}{Zenglin Xu}, {and} \bibinfo{person}{Michael~R. Lyu}.} \bibinfo{year}{2021}\natexlab{}.
\newblock \showarticletitle{CRaDLe: Deep code retrieval based on semantic Dependency Learning}.
\newblock \bibinfo{journal}{\emph{Neural Networks}}  \bibinfo{volume}{141} (\bibinfo{year}{2021}), \bibinfo{pages}{385--394}.
\newblock


\bibitem[Guan et~al\mbox{.}(2021)]%
        {DBLP:conf/acl/GuanMFLDH20}
\bibfield{author}{\bibinfo{person}{Jian Guan}, \bibinfo{person}{Xiaoxi Mao}, \bibinfo{person}{Changjie Fan}, \bibinfo{person}{Zitao Liu}, \bibinfo{person}{Wenbiao Ding}, {and} \bibinfo{person}{Minlie Huang}.} \bibinfo{year}{2021}\natexlab{}.
\newblock \showarticletitle{Long Text Generation by Modeling Sentence-Level and Discourse-Level Coherence}. In \bibinfo{booktitle}{\emph{Proceedings of the 59th Annual Meeting of the Association for Computational Linguistics and the 11th International Joint Conference on Natural Language Processing, {ACL/IJCNLP} 2021, (Volume 1: Long Papers), Virtual Event, August 1-6, 2021}}, \bibfield{editor}{\bibinfo{person}{Chengqing Zong}, \bibinfo{person}{Fei Xia}, \bibinfo{person}{Wenjie Li}, {and} \bibinfo{person}{Roberto Navigli}} (Eds.). \bibinfo{publisher}{Association for Computational Linguistics}, \bibinfo{pages}{6379--6393}.
\newblock


\bibitem[Guo et~al\mbox{.}(2022)]%
        {DBLP:conf/acl/GuoLDW0022}
\bibfield{author}{\bibinfo{person}{Daya Guo}, \bibinfo{person}{Shuai Lu}, \bibinfo{person}{Nan Duan}, \bibinfo{person}{Yanlin Wang}, \bibinfo{person}{Ming Zhou}, {and} \bibinfo{person}{Jian Yin}.} \bibinfo{year}{2022}\natexlab{}.
\newblock \showarticletitle{UniXcoder: Unified Cross-Modal Pre-training for Code Representation}. In \bibinfo{booktitle}{\emph{Proceedings of the 60th Annual Meeting of the Association for Computational Linguistics (Volume 1: Long Papers), {ACL} 2022, Dublin, Ireland, May 22-27, 2022}}, \bibfield{editor}{\bibinfo{person}{Smaranda Muresan}, \bibinfo{person}{Preslav Nakov}, {and} \bibinfo{person}{Aline Villavicencio}} (Eds.). \bibinfo{publisher}{Association for Computational Linguistics}, \bibinfo{pages}{7212--7225}.
\newblock


\bibitem[Guo et~al\mbox{.}(2021)]%
        {DBLP:conf/iclr/GuoRLFT0ZDSFTDC21}
\bibfield{author}{\bibinfo{person}{Daya Guo}, \bibinfo{person}{Shuo Ren}, \bibinfo{person}{Shuai Lu}, \bibinfo{person}{Zhangyin Feng}, \bibinfo{person}{Duyu Tang}, \bibinfo{person}{Shujie Liu}, \bibinfo{person}{Long Zhou}, \bibinfo{person}{Nan Duan}, \bibinfo{person}{Alexey Svyatkovskiy}, \bibinfo{person}{Shengyu Fu}, \bibinfo{person}{Michele Tufano}, \bibinfo{person}{Shao~Kun Deng}, \bibinfo{person}{Colin~B. Clement}, \bibinfo{person}{Dawn Drain}, \bibinfo{person}{Neel Sundaresan}, \bibinfo{person}{Jian Yin}, \bibinfo{person}{Daxin Jiang}, {and} \bibinfo{person}{Ming Zhou}.} \bibinfo{year}{2021}\natexlab{}.
\newblock \showarticletitle{GraphCodeBERT: Pre-training Code Representations with Data Flow}. In \bibinfo{booktitle}{\emph{9th International Conference on Learning Representations, {ICLR} 2021, Virtual Event, Austria, May 3-7, 2021}}. \bibinfo{publisher}{OpenReview.net}.
\newblock


\bibitem[Guo et~al\mbox{.}(2023)]%
        {DBLP:conf/icml/GuoXD0M23}
\bibfield{author}{\bibinfo{person}{Daya Guo}, \bibinfo{person}{Canwen Xu}, \bibinfo{person}{Nan Duan}, \bibinfo{person}{Jian Yin}, {and} \bibinfo{person}{Julian~J. McAuley}.} \bibinfo{year}{2023}\natexlab{}.
\newblock \showarticletitle{LongCoder: {A} Long-Range Pre-trained Language Model for Code Completion}. In \bibinfo{booktitle}{\emph{International Conference on Machine Learning, {ICML} 2023, 23-29 July 2023, Honolulu, Hawaii, {USA}}} \emph{(\bibinfo{series}{Proceedings of Machine Learning Research}, Vol.~\bibinfo{volume}{202})}, \bibfield{editor}{\bibinfo{person}{Andreas Krause}, \bibinfo{person}{Emma Brunskill}, \bibinfo{person}{Kyunghyun Cho}, \bibinfo{person}{Barbara Engelhardt}, \bibinfo{person}{Sivan Sabato}, {and} \bibinfo{person}{Jonathan Scarlett}} (Eds.). \bibinfo{publisher}{{PMLR}}, \bibinfo{pages}{12098--12107}.
\newblock


\bibitem[Hsieh et~al\mbox{.}(2023)]%
        {DBLP:conf/acl/HsiehLYNFRKLP23}
\bibfield{author}{\bibinfo{person}{Cheng{-}Yu Hsieh}, \bibinfo{person}{Chun{-}Liang Li}, \bibinfo{person}{Chih{-}Kuan Yeh}, \bibinfo{person}{Hootan Nakhost}, \bibinfo{person}{Yasuhisa Fujii}, \bibinfo{person}{Alex Ratner}, \bibinfo{person}{Ranjay Krishna}, \bibinfo{person}{Chen{-}Yu Lee}, {and} \bibinfo{person}{Tomas Pfister}.} \bibinfo{year}{2023}\natexlab{}.
\newblock \showarticletitle{Distilling Step-by-Step! Outperforming Larger Language Models with Less Training Data and Smaller Model Sizes}. In \bibinfo{booktitle}{\emph{Findings of the Association for Computational Linguistics: {ACL} 2023, Toronto, Canada, July 9-14, 2023}}, \bibfield{editor}{\bibinfo{person}{Anna Rogers}, \bibinfo{person}{Jordan~L. Boyd{-}Graber}, {and} \bibinfo{person}{Naoaki Okazaki}} (Eds.). \bibinfo{publisher}{Association for Computational Linguistics}, \bibinfo{pages}{8003--8017}.
\newblock


\bibitem[Hu et~al\mbox{.}(2018)]%
        {DBLP:conf/iwpc/HuLXLJ18}
\bibfield{author}{\bibinfo{person}{Xing Hu}, \bibinfo{person}{Ge Li}, \bibinfo{person}{Xin Xia}, \bibinfo{person}{David Lo}, {and} \bibinfo{person}{Zhi Jin}.} \bibinfo{year}{2018}\natexlab{}.
\newblock \showarticletitle{Deep code comment generation}. In \bibinfo{booktitle}{\emph{Proceedings of the 26th Conference on Program Comprehension, {ICPC} 2018, Gothenburg, Sweden, May 27-28, 2018}}, \bibfield{editor}{\bibinfo{person}{Foutse Khomh}, \bibinfo{person}{Chanchal~K. Roy}, {and} \bibinfo{person}{Janet Siegmund}} (Eds.). \bibinfo{publisher}{{ACM}}, \bibinfo{pages}{200--210}.
\newblock


\bibitem[Hub(2023)]%
        {HuggingFace}
\bibfield{author}{\bibinfo{person}{Huggingface Hub}.} \bibinfo{year}{2023}\natexlab{}.
\newblock
\newblock
\urldef\tempurl%
\url{https://huggingface.co/}
\showURL{%
\tempurl}


\bibitem[Kang et~al\mbox{.}(2021)]%
        {DBLP:conf/emnlp/KangW00Y21}
\bibfield{author}{\bibinfo{person}{Yuning Kang}, \bibinfo{person}{Zan Wang}, \bibinfo{person}{Hongyu Zhang}, \bibinfo{person}{Junjie Chen}, {and} \bibinfo{person}{Hanmo You}.} \bibinfo{year}{2021}\natexlab{}.
\newblock \showarticletitle{APIRecX: Cross-Library {API} Recommendation via Pre-Trained Language Model}. In \bibinfo{booktitle}{\emph{Proceedings of the 2021 Conference on Empirical Methods in Natural Language Processing, {EMNLP} 2021, Virtual Event / Punta Cana, Dominican Republic, 7-11 November, 2021}}, \bibfield{editor}{\bibinfo{person}{Marie{-}Francine Moens}, \bibinfo{person}{Xuanjing Huang}, \bibinfo{person}{Lucia Specia}, {and} \bibinfo{person}{Scott~Wen{-}tau Yih}} (Eds.). \bibinfo{publisher}{Association for Computational Linguistics}, \bibinfo{pages}{3425--3436}.
\newblock


\bibitem[Khare et~al\mbox{.}(2023)]%
        {DBLP:journals/corr/abs-2311-16169}
\bibfield{author}{\bibinfo{person}{Avishree Khare}, \bibinfo{person}{Saikat Dutta}, \bibinfo{person}{Ziyang Li}, \bibinfo{person}{Alaia Solko{-}Breslin}, \bibinfo{person}{Rajeev Alur}, {and} \bibinfo{person}{Mayur Naik}.} \bibinfo{year}{2023}\natexlab{}.
\newblock \showarticletitle{Understanding the Effectiveness of Large Language Models in Detecting Security Vulnerabilities}.
\newblock \bibinfo{journal}{\emph{CoRR}}  \bibinfo{volume}{abs/2311.16169} (\bibinfo{year}{2023}).
\newblock


\bibitem[Kim et~al\mbox{.}(2021)]%
        {DBLP:conf/icse/KimZT021}
\bibfield{author}{\bibinfo{person}{Seohyun Kim}, \bibinfo{person}{Jinman Zhao}, \bibinfo{person}{Yuchi Tian}, {and} \bibinfo{person}{Satish Chandra}.} \bibinfo{year}{2021}\natexlab{}.
\newblock \showarticletitle{Code Prediction by Feeding Trees to Transformers}. In \bibinfo{booktitle}{\emph{43rd {IEEE/ACM} International Conference on Software Engineering, {ICSE} 2021, Madrid, Spain, 22-30 May 2021}}. \bibinfo{publisher}{{IEEE}}, \bibinfo{pages}{150--162}.
\newblock


\bibitem[Kingma and Ba(2015)]%
        {DBLP:journals/corr/KingmaB14}
\bibfield{author}{\bibinfo{person}{Diederik~P. Kingma} {and} \bibinfo{person}{Jimmy Ba}.} \bibinfo{year}{2015}\natexlab{}.
\newblock \showarticletitle{Adam: {A} Method for Stochastic Optimization}. In \bibinfo{booktitle}{\emph{3rd International Conference on Learning Representations, {ICLR} 2015, San Diego, CA, USA, May 7-9, 2015, Conference Track Proceedings}}, \bibfield{editor}{\bibinfo{person}{Yoshua Bengio} {and} \bibinfo{person}{Yann LeCun}} (Eds.).
\newblock


\bibitem[Kula et~al\mbox{.}(2018)]%
        {DBLP:journals/ese/KulaGOII18}
\bibfield{author}{\bibinfo{person}{Raula~Gaikovina Kula}, \bibinfo{person}{Daniel~M. Germ{\'{a}}n}, \bibinfo{person}{Ali Ouni}, \bibinfo{person}{Takashi Ishio}, {and} \bibinfo{person}{Katsuro Inoue}.} \bibinfo{year}{2018}\natexlab{}.
\newblock \showarticletitle{Do developers update their library dependencies? - An empirical study on the impact of security advisories on library migration}.
\newblock \bibinfo{journal}{\emph{Empir. Softw. Eng.}} \bibinfo{volume}{23}, \bibinfo{number}{1} (\bibinfo{year}{2018}), \bibinfo{pages}{384--417}.
\newblock


\bibitem[Li et~al\mbox{.}(2021)]%
        {DBLP:conf/sigsoft/Li0N21}
\bibfield{author}{\bibinfo{person}{Yi Li}, \bibinfo{person}{Shaohua Wang}, {and} \bibinfo{person}{Tien~N. Nguyen}.} \bibinfo{year}{2021}\natexlab{}.
\newblock \showarticletitle{Vulnerability detection with fine-grained interpretations}. In \bibinfo{booktitle}{\emph{{ESEC/FSE} '21: 29th {ACM} Joint European Software Engineering Conference and Symposium on the Foundations of Software Engineering, Athens, Greece, August 23-28, 2021}}, \bibfield{editor}{\bibinfo{person}{Diomidis Spinellis}, \bibinfo{person}{Georgios Gousios}, \bibinfo{person}{Marsha Chechik}, {and} \bibinfo{person}{Massimiliano~Di Penta}} (Eds.). \bibinfo{publisher}{{ACM}}, \bibinfo{pages}{292--303}.
\newblock


\bibitem[Li et~al\mbox{.}(2023)]%
        {ptm4api}
\bibfield{author}{\bibinfo{person}{Zhihao Li}, \bibinfo{person}{Chuanyi Li}, \bibinfo{person}{Ze Tang}, \bibinfo{person}{Wanhong Huang}, \bibinfo{person}{Jidong Ge}, \bibinfo{person}{Bin Luo}, \bibinfo{person}{Vincent Ng}, \bibinfo{person}{Ting Wang}, \bibinfo{person}{Yucheng Hu}, {and} \bibinfo{person}{Xiaopeng Zhang}.} \bibinfo{year}{2023}\natexlab{}.
\newblock \showarticletitle{PTM-APIRec: Leveraging Pre-trained Models of Source Code in API Recommendation}.
\newblock \bibinfo{journal}{\emph{ACM Transactions on Software Engineering and Methodology}} (\bibinfo{year}{2023}).
\newblock


\bibitem[Li et~al\mbox{.}(2022)]%
        {DBLP:journals/tdsc/0027ZX0ZC22}
\bibfield{author}{\bibinfo{person}{Zhen Li}, \bibinfo{person}{Deqing Zou}, \bibinfo{person}{Shouhuai Xu}, \bibinfo{person}{Hai Jin}, \bibinfo{person}{Yawei Zhu}, {and} \bibinfo{person}{Zhaoxuan Chen}.} \bibinfo{year}{2022}\natexlab{}.
\newblock \showarticletitle{SySeVR: {A} Framework for Using Deep Learning to Detect Software Vulnerabilities}.
\newblock \bibinfo{journal}{\emph{{IEEE} Trans. Dependable Secur. Comput.}} \bibinfo{volume}{19}, \bibinfo{number}{4} (\bibinfo{year}{2022}), \bibinfo{pages}{2244--2258}.
\newblock


\bibitem[Li et~al\mbox{.}(2018)]%
        {DBLP:conf/ndss/LiZXO0WDZ18}
\bibfield{author}{\bibinfo{person}{Zhen Li}, \bibinfo{person}{Deqing Zou}, \bibinfo{person}{Shouhuai Xu}, \bibinfo{person}{Xinyu Ou}, \bibinfo{person}{Hai Jin}, \bibinfo{person}{Sujuan Wang}, \bibinfo{person}{Zhijun Deng}, {and} \bibinfo{person}{Yuyi Zhong}.} \bibinfo{year}{2018}\natexlab{}.
\newblock \showarticletitle{VulDeePecker: {A} Deep Learning-Based System for Vulnerability Detection}. In \bibinfo{booktitle}{\emph{25th Annual Network and Distributed System Security Symposium, {NDSS} 2018, San Diego, California, USA, February 18-21, 2018}}. \bibinfo{publisher}{The Internet Society}.
\newblock


\bibitem[Liu et~al\mbox{.}(2019)]%
        {DBLP:journals/corr/abs-1907-11692}
\bibfield{author}{\bibinfo{person}{Yinhan Liu}, \bibinfo{person}{Myle Ott}, \bibinfo{person}{Naman Goyal}, \bibinfo{person}{Jingfei Du}, \bibinfo{person}{Mandar Joshi}, \bibinfo{person}{Danqi Chen}, \bibinfo{person}{Omer Levy}, \bibinfo{person}{Mike Lewis}, \bibinfo{person}{Luke Zettlemoyer}, {and} \bibinfo{person}{Veselin Stoyanov}.} \bibinfo{year}{2019}\natexlab{}.
\newblock \showarticletitle{RoBERTa: {A} Robustly Optimized {BERT} Pretraining Approach}.
\newblock \bibinfo{journal}{\emph{CoRR}}  \bibinfo{volume}{abs/1907.11692} (\bibinfo{year}{2019}).
\newblock


\bibitem[Lu et~al\mbox{.}(2021)]%
        {DBLP:conf/nips/LuGRHSBCDJTLZSZ21}
\bibfield{author}{\bibinfo{person}{Shuai Lu}, \bibinfo{person}{Daya Guo}, \bibinfo{person}{Shuo Ren}, \bibinfo{person}{Junjie Huang}, \bibinfo{person}{Alexey Svyatkovskiy}, \bibinfo{person}{Ambrosio Blanco}, \bibinfo{person}{Colin~B. Clement}, \bibinfo{person}{Dawn Drain}, \bibinfo{person}{Daxin Jiang}, \bibinfo{person}{Duyu Tang}, \bibinfo{person}{Ge Li}, \bibinfo{person}{Lidong Zhou}, \bibinfo{person}{Linjun Shou}, \bibinfo{person}{Long Zhou}, \bibinfo{person}{Michele Tufano}, \bibinfo{person}{Ming Gong}, \bibinfo{person}{Ming Zhou}, \bibinfo{person}{Nan Duan}, \bibinfo{person}{Neel Sundaresan}, \bibinfo{person}{Shao~Kun Deng}, \bibinfo{person}{Shengyu Fu}, {and} \bibinfo{person}{Shujie Liu}.} \bibinfo{year}{2021}\natexlab{}.
\newblock \showarticletitle{CodeXGLUE: {A} Machine Learning Benchmark Dataset for Code Understanding and Generation}. In \bibinfo{booktitle}{\emph{Proceedings of the Neural Information Processing Systems Track on Datasets and Benchmarks 1, NeurIPS Datasets and Benchmarks 2021, December 2021, virtual}}, \bibfield{editor}{\bibinfo{person}{Joaquin Vanschoren} {and} \bibinfo{person}{Sai{-}Kit Yeung}} (Eds.).
\newblock


\bibitem[McIntosh et~al\mbox{.}(2011)]%
        {DBLP:conf/icse/McIntoshANKH11}
\bibfield{author}{\bibinfo{person}{Shane McIntosh}, \bibinfo{person}{Bram Adams}, \bibinfo{person}{Thanh H.~D. Nguyen}, \bibinfo{person}{Yasutaka Kamei}, {and} \bibinfo{person}{Ahmed~E. Hassan}.} \bibinfo{year}{2011}\natexlab{}.
\newblock \showarticletitle{An empirical study of build maintenance effort}. In \bibinfo{booktitle}{\emph{Proceedings of the 33rd International Conference on Software Engineering, {ICSE} 2011, Waikiki, Honolulu , HI, USA, May 21-28, 2011}}, \bibfield{editor}{\bibinfo{person}{Richard~N. Taylor}, \bibinfo{person}{Harald~C. Gall}, {and} \bibinfo{person}{Nenad Medvidovic}} (Eds.). \bibinfo{publisher}{{ACM}}, \bibinfo{pages}{141--150}.
\newblock


\bibitem[McIntosh et~al\mbox{.}(2016)]%
        {DBLP:journals/ese/McIntoshKAH16}
\bibfield{author}{\bibinfo{person}{Shane McIntosh}, \bibinfo{person}{Yasutaka Kamei}, \bibinfo{person}{Bram Adams}, {and} \bibinfo{person}{Ahmed~E. Hassan}.} \bibinfo{year}{2016}\natexlab{}.
\newblock \showarticletitle{An empirical study of the impact of modern code review practices on software quality}.
\newblock \bibinfo{journal}{\emph{Empir. Softw. Eng.}} \bibinfo{volume}{21}, \bibinfo{number}{5} (\bibinfo{year}{2016}), \bibinfo{pages}{2146--2189}.
\newblock


\bibitem[Microsoft(2023)]%
        {Github}
\bibfield{author}{\bibinfo{person}{Microsoft}.} \bibinfo{year}{2023}\natexlab{}.
\newblock \bibinfo{title}{Github website.}
\newblock
\newblock
\urldef\tempurl%
\url{https://github.com/}
\showURL{%
\tempurl}


\bibitem[Nguyen and Nguyen(2015)]%
        {DBLP:conf/icse/NguyenN15}
\bibfield{author}{\bibinfo{person}{Anh~Tuan Nguyen} {and} \bibinfo{person}{Tien~N. Nguyen}.} \bibinfo{year}{2015}\natexlab{}.
\newblock \showarticletitle{Graph-Based Statistical Language Model for Code}. In \bibinfo{booktitle}{\emph{37th {IEEE/ACM} International Conference on Software Engineering, {ICSE} 2015, Florence, Italy, May 16-24, 2015, Volume 1}}, \bibfield{editor}{\bibinfo{person}{Antonia Bertolino}, \bibinfo{person}{Gerardo Canfora}, {and} \bibinfo{person}{Sebastian~G. Elbaum}} (Eds.). \bibinfo{publisher}{{IEEE} Computer Society}, \bibinfo{pages}{858--868}.
\newblock


\bibitem[Nijkamp et~al\mbox{.}(2023a)]%
        {Nijkamp2023codegen2}
\bibfield{author}{\bibinfo{person}{Erik Nijkamp}, \bibinfo{person}{Hiroaki Hayashi}, \bibinfo{person}{Caiming Xiong}, \bibinfo{person}{Silvio Savarese}, {and} \bibinfo{person}{Yingbo Zhou}.} \bibinfo{year}{2023}\natexlab{a}.
\newblock \showarticletitle{CodeGen2: Lessons for Training LLMs on Programming and Natural Languages}.
\newblock \bibinfo{journal}{\emph{arXiv preprint}} (\bibinfo{year}{2023}).
\newblock


\bibitem[Nijkamp et~al\mbox{.}(2023b)]%
        {DBLP:journals/corr/abs-2305-02309}
\bibfield{author}{\bibinfo{person}{Erik Nijkamp}, \bibinfo{person}{Hiroaki Hayashi}, \bibinfo{person}{Caiming Xiong}, \bibinfo{person}{Silvio Savarese}, {and} \bibinfo{person}{Yingbo Zhou}.} \bibinfo{year}{2023}\natexlab{b}.
\newblock \showarticletitle{CodeGen2: Lessons for Training LLMs on Programming and Natural Languages}.
\newblock \bibinfo{journal}{\emph{CoRR}}  \bibinfo{volume}{abs/2305.02309} (\bibinfo{year}{2023}).
\newblock


\bibitem[Nijkamp et~al\mbox{.}(2023c)]%
        {DBLP:conf/iclr/NijkampPHTWZSX23}
\bibfield{author}{\bibinfo{person}{Erik Nijkamp}, \bibinfo{person}{Bo Pang}, \bibinfo{person}{Hiroaki Hayashi}, \bibinfo{person}{Lifu Tu}, \bibinfo{person}{Huan Wang}, \bibinfo{person}{Yingbo Zhou}, \bibinfo{person}{Silvio Savarese}, {and} \bibinfo{person}{Caiming Xiong}.} \bibinfo{year}{2023}\natexlab{c}.
\newblock \showarticletitle{CodeGen: An Open Large Language Model for Code with Multi-Turn Program Synthesis}. In \bibinfo{booktitle}{\emph{The Eleventh International Conference on Learning Representations, {ICLR} 2023, Kigali, Rwanda, May 1-5, 2023}}. \bibinfo{publisher}{OpenReview.net}.
\newblock


\bibitem[Nikitopoulos et~al\mbox{.}(2021)]%
        {DBLP:conf/sigsoft/NikitopoulosDLM21}
\bibfield{author}{\bibinfo{person}{Georgios Nikitopoulos}, \bibinfo{person}{Konstantina Dritsa}, \bibinfo{person}{Panos Louridas}, {and} \bibinfo{person}{Dimitris Mitropoulos}.} \bibinfo{year}{2021}\natexlab{}.
\newblock \showarticletitle{CrossVul: a cross-language vulnerability dataset with commit data}. In \bibinfo{booktitle}{\emph{{ESEC/FSE} '21: 29th {ACM} Joint European Software Engineering Conference and Symposium on the Foundations of Software Engineering, Athens, Greece, August 23-28, 2021}}, \bibfield{editor}{\bibinfo{person}{Diomidis Spinellis}, \bibinfo{person}{Georgios Gousios}, \bibinfo{person}{Marsha Chechik}, {and} \bibinfo{person}{Massimiliano~Di Penta}} (Eds.). \bibinfo{publisher}{{ACM}}, \bibinfo{pages}{1565--1569}.
\newblock


\bibitem[OpenAI(2023)]%
        {gpt4}
\bibfield{author}{\bibinfo{person}{OpenAI}.} \bibinfo{year}{2023}\natexlab{}.
\newblock \showarticletitle{{GPT-4} Technical Report}.
\newblock \bibinfo{journal}{\emph{CoRR}}  \bibinfo{volume}{abs/2303.08774} (\bibinfo{year}{2023}).
\newblock


\bibitem[Padioleau et~al\mbox{.}(2009)]%
        {DBLP:conf/icse/PadioleauTZ09}
\bibfield{author}{\bibinfo{person}{Yoann Padioleau}, \bibinfo{person}{Lin Tan}, {and} \bibinfo{person}{Yuanyuan Zhou}.} \bibinfo{year}{2009}\natexlab{}.
\newblock \showarticletitle{Listening to programmers - Taxonomies and characteristics of comments in operating system code}. In \bibinfo{booktitle}{\emph{31st International Conference on Software Engineering, {ICSE} 2009, May 16-24, 2009, Vancouver, Canada, Proceedings}}. \bibinfo{publisher}{{IEEE}}, \bibinfo{pages}{331--341}.
\newblock


\bibitem[Peng et~al\mbox{.}(2023)]%
        {DBLP:journals/tse/PengLGLWGL23}
\bibfield{author}{\bibinfo{person}{Yun Peng}, \bibinfo{person}{Shuqing Li}, \bibinfo{person}{Wenwei Gu}, \bibinfo{person}{Yichen Li}, \bibinfo{person}{Wenxuan Wang}, \bibinfo{person}{Cuiyun Gao}, {and} \bibinfo{person}{Michael~R. Lyu}.} \bibinfo{year}{2023}\natexlab{}.
\newblock \showarticletitle{Revisiting, Benchmarking and Exploring {API} Recommendation: How Far Are We?}
\newblock \bibinfo{journal}{\emph{{IEEE} Trans. Software Eng.}} \bibinfo{volume}{49}, \bibinfo{number}{4} (\bibinfo{year}{2023}), \bibinfo{pages}{1876--1897}.
\newblock


\bibitem[Radford et~al\mbox{.}(2018)]%
        {gpt1}
\bibfield{author}{\bibinfo{person}{Alec Radford}, \bibinfo{person}{Karthik Narasimhan}, \bibinfo{person}{Tim Salimans}, \bibinfo{person}{Ilya Sutskever}, {et~al\mbox{.}}} \bibinfo{year}{2018}\natexlab{}.
\newblock \showarticletitle{Improving language understanding by generative pre-training}.
\newblock  (\bibinfo{year}{2018}).
\newblock


\bibitem[Radford et~al\mbox{.}(2019)]%
        {gpt2}
\bibfield{author}{\bibinfo{person}{Alec Radford}, \bibinfo{person}{Jeffrey Wu}, \bibinfo{person}{Rewon Child}, \bibinfo{person}{David Luan}, \bibinfo{person}{Dario Amodei}, \bibinfo{person}{Ilya Sutskever}, {et~al\mbox{.}}} \bibinfo{year}{2019}\natexlab{}.
\newblock \showarticletitle{Language models are unsupervised multitask learners}.
\newblock \bibinfo{journal}{\emph{OpenAI blog}} \bibinfo{volume}{1}, \bibinfo{number}{8} (\bibinfo{year}{2019}), \bibinfo{pages}{9}.
\newblock


\bibitem[Raffel et~al\mbox{.}(2020a)]%
        {DBLP:journals/jmlr/RaffelSRLNMZLL20}
\bibfield{author}{\bibinfo{person}{Colin Raffel}, \bibinfo{person}{Noam Shazeer}, \bibinfo{person}{Adam Roberts}, \bibinfo{person}{Katherine Lee}, \bibinfo{person}{Sharan Narang}, \bibinfo{person}{Michael Matena}, \bibinfo{person}{Yanqi Zhou}, \bibinfo{person}{Wei Li}, {and} \bibinfo{person}{Peter~J. Liu}.} \bibinfo{year}{2020}\natexlab{a}.
\newblock \showarticletitle{Exploring the Limits of Transfer Learning with a Unified Text-to-Text Transformer}.
\newblock \bibinfo{journal}{\emph{J. Mach. Learn. Res.}}  \bibinfo{volume}{21} (\bibinfo{year}{2020}), \bibinfo{pages}{140:1--140:67}.
\newblock


\bibitem[Raffel et~al\mbox{.}(2020b)]%
        {T5}
\bibfield{author}{\bibinfo{person}{Colin Raffel}, \bibinfo{person}{Noam Shazeer}, \bibinfo{person}{Adam Roberts}, \bibinfo{person}{Katherine Lee}, \bibinfo{person}{Sharan Narang}, \bibinfo{person}{Michael Matena}, \bibinfo{person}{Yanqi Zhou}, \bibinfo{person}{Wei Li}, {and} \bibinfo{person}{Peter~J. Liu}.} \bibinfo{year}{2020}\natexlab{b}.
\newblock \showarticletitle{Exploring the Limits of Transfer Learning with a Unified Text-to-Text Transformer}.
\newblock \bibinfo{journal}{\emph{J. Mach. Learn. Res.}}  \bibinfo{volume}{21} (\bibinfo{year}{2020}), \bibinfo{pages}{140:1--140:67}.
\newblock


\bibitem[Russell et~al\mbox{.}(2018)]%
        {DBLP:conf/icmla/RussellKHLHOEM18}
\bibfield{author}{\bibinfo{person}{Rebecca~L. Russell}, \bibinfo{person}{Louis~Y. Kim}, \bibinfo{person}{Lei~H. Hamilton}, \bibinfo{person}{Tomo Lazovich}, \bibinfo{person}{Jacob Harer}, \bibinfo{person}{Onur Ozdemir}, \bibinfo{person}{Paul~M. Ellingwood}, {and} \bibinfo{person}{Marc~W. McConley}.} \bibinfo{year}{2018}\natexlab{}.
\newblock \showarticletitle{Automated Vulnerability Detection in Source Code Using Deep Representation Learning}. In \bibinfo{booktitle}{\emph{17th {IEEE} International Conference on Machine Learning and Applications, {ICMLA} 2018, Orlando, FL, USA, December 17-20, 2018}}, \bibfield{editor}{\bibinfo{person}{M.~Arif Wani}, \bibinfo{person}{Mehmed~M. Kantardzic}, \bibinfo{person}{Moamar~Sayed Mouchaweh}, \bibinfo{person}{Jo{\~{a}}o Gama}, {and} \bibinfo{person}{Edwin Lughofer}} (Eds.). \bibinfo{publisher}{{IEEE}}, \bibinfo{pages}{757--762}.
\newblock


\bibitem[Touvron et~al\mbox{.}(2023)]%
        {llama2}
\bibfield{author}{\bibinfo{person}{Hugo Touvron}, \bibinfo{person}{Louis Martin}, \bibinfo{person}{Kevin Stone}, \bibinfo{person}{Peter Albert}, \bibinfo{person}{Amjad Almahairi}, \bibinfo{person}{Yasmine Babaei}, \bibinfo{person}{Nikolay Bashlykov}, \bibinfo{person}{Soumya Batra}, \bibinfo{person}{Prajjwal Bhargava}, \bibinfo{person}{Shruti Bhosale}, \bibinfo{person}{Dan Bikel}, \bibinfo{person}{Lukas Blecher}, \bibinfo{person}{Cristian Canton{-}Ferrer}, \bibinfo{person}{Moya Chen}, \bibinfo{person}{Guillem Cucurull}, \bibinfo{person}{David Esiobu}, \bibinfo{person}{Jude Fernandes}, \bibinfo{person}{Jeremy Fu}, \bibinfo{person}{Wenyin Fu}, \bibinfo{person}{Brian Fuller}, \bibinfo{person}{Cynthia Gao}, \bibinfo{person}{Vedanuj Goswami}, \bibinfo{person}{Naman Goyal}, \bibinfo{person}{Anthony Hartshorn}, \bibinfo{person}{Saghar Hosseini}, \bibinfo{person}{Rui Hou}, \bibinfo{person}{Hakan Inan}, \bibinfo{person}{Marcin Kardas}, \bibinfo{person}{Viktor Kerkez}, \bibinfo{person}{Madian Khabsa},
  \bibinfo{person}{Isabel Kloumann}, \bibinfo{person}{Artem Korenev}, \bibinfo{person}{Punit~Singh Koura}, \bibinfo{person}{Marie{-}Anne Lachaux}, \bibinfo{person}{Thibaut Lavril}, \bibinfo{person}{Jenya Lee}, \bibinfo{person}{Diana Liskovich}, \bibinfo{person}{Yinghai Lu}, \bibinfo{person}{Yuning Mao}, \bibinfo{person}{Xavier Martinet}, \bibinfo{person}{Todor Mihaylov}, \bibinfo{person}{Pushkar Mishra}, \bibinfo{person}{Igor Molybog}, \bibinfo{person}{Yixin Nie}, \bibinfo{person}{Andrew Poulton}, \bibinfo{person}{Jeremy Reizenstein}, \bibinfo{person}{Rashi Rungta}, \bibinfo{person}{Kalyan Saladi}, \bibinfo{person}{Alan Schelten}, \bibinfo{person}{Ruan Silva}, \bibinfo{person}{Eric~Michael Smith}, \bibinfo{person}{Ranjan Subramanian}, \bibinfo{person}{Xiaoqing~Ellen Tan}, \bibinfo{person}{Binh Tang}, \bibinfo{person}{Ross Taylor}, \bibinfo{person}{Adina Williams}, \bibinfo{person}{Jian~Xiang Kuan}, \bibinfo{person}{Puxin Xu}, \bibinfo{person}{Zheng Yan}, \bibinfo{person}{Iliyan Zarov}, \bibinfo{person}{Yuchen
  Zhang}, \bibinfo{person}{Angela Fan}, \bibinfo{person}{Melanie Kambadur}, \bibinfo{person}{Sharan Narang}, \bibinfo{person}{Aur{\'{e}}lien Rodriguez}, \bibinfo{person}{Robert Stojnic}, \bibinfo{person}{Sergey Edunov}, {and} \bibinfo{person}{Thomas Scialom}.} \bibinfo{year}{2023}\natexlab{}.
\newblock \showarticletitle{Llama 2: Open Foundation and Fine-Tuned Chat Models}.
\newblock \bibinfo{journal}{\emph{CoRR}}  \bibinfo{volume}{abs/2307.09288} (\bibinfo{year}{2023}).
\newblock


\bibitem[Tree{-}sitter(2023)]%
        {tree-sitter}
\bibfield{author}{\bibinfo{person}{Tree{-}sitter}.} \bibinfo{year}{2023}\natexlab{}.
\newblock
\newblock
\urldef\tempurl%
\url{https://github.com/tree-sitter/tree-sitter}
\showURL{%
\tempurl}


\bibitem[Wang et~al\mbox{.}(2022)]%
        {DBLP:conf/sigsoft/WangYGP0L22}
\bibfield{author}{\bibinfo{person}{Chaozheng Wang}, \bibinfo{person}{Yuanhang Yang}, \bibinfo{person}{Cuiyun Gao}, \bibinfo{person}{Yun Peng}, \bibinfo{person}{Hongyu Zhang}, {and} \bibinfo{person}{Michael~R. Lyu}.} \bibinfo{year}{2022}\natexlab{}.
\newblock \showarticletitle{No more fine-tuning? an experimental evaluation of prompt tuning in code intelligence}. In \bibinfo{booktitle}{\emph{Proceedings of the 30th {ACM} Joint European Software Engineering Conference and Symposium on the Foundations of Software Engineering, {ESEC/FSE} 2022, Singapore, Singapore, November 14-18, 2022}}, \bibfield{editor}{\bibinfo{person}{Abhik Roychoudhury}, \bibinfo{person}{Cristian Cadar}, {and} \bibinfo{person}{Miryung Kim}} (Eds.). \bibinfo{publisher}{{ACM}}, \bibinfo{pages}{382--394}.
\newblock


\bibitem[Wang et~al\mbox{.}(2021)]%
        {DBLP:conf/emnlp/0034WJH21}
\bibfield{author}{\bibinfo{person}{Yue Wang}, \bibinfo{person}{Weishi Wang}, \bibinfo{person}{Shafiq~R. Joty}, {and} \bibinfo{person}{Steven C.~H. Hoi}.} \bibinfo{year}{2021}\natexlab{}.
\newblock \showarticletitle{CodeT5: Identifier-aware Unified Pre-trained Encoder-Decoder Models for Code Understanding and Generation}. In \bibinfo{booktitle}{\emph{Proceedings of the 2021 Conference on Empirical Methods in Natural Language Processing, {EMNLP} 2021, Virtual Event / Punta Cana, Dominican Republic, 7-11 November, 2021}}, \bibfield{editor}{\bibinfo{person}{Marie{-}Francine Moens}, \bibinfo{person}{Xuanjing Huang}, \bibinfo{person}{Lucia Specia}, {and} \bibinfo{person}{Scott~Wen{-}tau Yih}} (Eds.). \bibinfo{publisher}{Association for Computational Linguistics}, \bibinfo{pages}{8696--8708}.
\newblock


\bibitem[Wei et~al\mbox{.}(2022)]%
        {DBLP:conf/icse/WeiHH0022}
\bibfield{author}{\bibinfo{person}{Moshi Wei}, \bibinfo{person}{Nima~Shiri Harzevili}, \bibinfo{person}{Yuchao Huang}, \bibinfo{person}{Junjie Wang}, {and} \bibinfo{person}{Song Wang}.} \bibinfo{year}{2022}\natexlab{}.
\newblock \showarticletitle{{CLEAR:} Contrastive Learning for {API} Recommendation}. In \bibinfo{booktitle}{\emph{44th {IEEE/ACM} 44th International Conference on Software Engineering, {ICSE} 2022, Pittsburgh, PA, USA, May 25-27, 2022}}. \bibinfo{publisher}{{ACM}}, \bibinfo{pages}{376--387}.
\newblock


\bibitem[Wei et~al\mbox{.}(2023)]%
        {DBLP:conf/sigsoft/0003X023}
\bibfield{author}{\bibinfo{person}{Yuxiang Wei}, \bibinfo{person}{Chunqiu~Steven Xia}, {and} \bibinfo{person}{Lingming Zhang}.} \bibinfo{year}{2023}\natexlab{}.
\newblock \showarticletitle{Copiloting the Copilots: Fusing Large Language Models with Completion Engines for Automated Program Repair}. In \bibinfo{booktitle}{\emph{Proceedings of the 31st {ACM} Joint European Software Engineering Conference and Symposium on the Foundations of Software Engineering, {ESEC/FSE} 2023, San Francisco, CA, USA, December 3-9, 2023}}, \bibfield{editor}{\bibinfo{person}{Satish Chandra}, \bibinfo{person}{Kelly Blincoe}, {and} \bibinfo{person}{Paolo Tonella}} (Eds.). \bibinfo{publisher}{{ACM}}, \bibinfo{pages}{172--184}.
\newblock


\bibitem[Wen et~al\mbox{.}(2023a)]%
        {DBLP:conf/icse/WenCGZZL23}
\bibfield{author}{\bibinfo{person}{Xin{-}Cheng Wen}, \bibinfo{person}{Yupan Chen}, \bibinfo{person}{Cuiyun Gao}, \bibinfo{person}{Hongyu Zhang}, \bibinfo{person}{Jie~M. Zhang}, {and} \bibinfo{person}{Qing Liao}.} \bibinfo{year}{2023}\natexlab{a}.
\newblock \showarticletitle{Vulnerability Detection with Graph Simplification and Enhanced Graph Representation Learning}. In \bibinfo{booktitle}{\emph{45th {IEEE/ACM} International Conference on Software Engineering, {ICSE} 2023, Melbourne, Australia, May 14-20, 2023}}. \bibinfo{publisher}{{IEEE}}, \bibinfo{pages}{2275--2286}.
\newblock


\bibitem[Wen et~al\mbox{.}(2023b)]%
        {DBLP:conf/kbse/WenWGWLG23}
\bibfield{author}{\bibinfo{person}{Xin{-}Cheng Wen}, \bibinfo{person}{Xinchen Wang}, \bibinfo{person}{Cuiyun Gao}, \bibinfo{person}{Shaohua Wang}, \bibinfo{person}{Yang Liu}, {and} \bibinfo{person}{Zhaoquan Gu}.} \bibinfo{year}{2023}\natexlab{b}.
\newblock \showarticletitle{When Less is Enough: Positive and Unlabeled Learning Model for Vulnerability Detection}. In \bibinfo{booktitle}{\emph{38th {IEEE/ACM} International Conference on Automated Software Engineering, {ASE} 2023, Luxembourg, September 11-15, 2023}}. \bibinfo{publisher}{{IEEE}}, \bibinfo{pages}{345--357}.
\newblock


\bibitem[Wilcoxon(1992)]%
        {wilcoxon1992individual}
\bibfield{author}{\bibinfo{person}{Frank Wilcoxon}.} \bibinfo{year}{1992}\natexlab{}.
\newblock \showarticletitle{Individual comparisons by ranking methods}.
\newblock In \bibinfo{booktitle}{\emph{Breakthroughs in statistics}}. \bibinfo{publisher}{Springer}, \bibinfo{pages}{196--202}.
\newblock


\bibitem[Zhang et~al\mbox{.}(2019)]%
        {astnn}
\bibfield{author}{\bibinfo{person}{Jian Zhang}, \bibinfo{person}{Xu Wang}, \bibinfo{person}{Hongyu Zhang}, \bibinfo{person}{Hailong Sun}, \bibinfo{person}{Kaixuan Wang}, {and} \bibinfo{person}{Xudong Liu}.} \bibinfo{year}{2019}\natexlab{}.
\newblock \showarticletitle{A novel neural source code representation based on abstract syntax tree}. In \bibinfo{booktitle}{\emph{Proceedings of the 41st International Conference on Software Engineering, {ICSE} 2019, Montreal, QC, Canada, May 25-31, 2019}}, \bibfield{editor}{\bibinfo{person}{Joanne~M. Atlee}, \bibinfo{person}{Tevfik Bultan}, {and} \bibinfo{person}{Jon Whittle}} (Eds.). \bibinfo{publisher}{{IEEE} / {ACM}}, \bibinfo{pages}{783--794}.
\newblock


\bibitem[Zheng et~al\mbox{.}(2023)]%
        {DBLP:journals/corr/abs-2303-17568}
\bibfield{author}{\bibinfo{person}{Qinkai Zheng}, \bibinfo{person}{Xiao Xia}, \bibinfo{person}{Xu Zou}, \bibinfo{person}{Yuxiao Dong}, \bibinfo{person}{Shan Wang}, \bibinfo{person}{Yufei Xue}, \bibinfo{person}{Zihan Wang}, \bibinfo{person}{Lei Shen}, \bibinfo{person}{Andi Wang}, \bibinfo{person}{Yang Li}, \bibinfo{person}{Teng Su}, \bibinfo{person}{Zhilin Yang}, {and} \bibinfo{person}{Jie Tang}.} \bibinfo{year}{2023}\natexlab{}.
\newblock \showarticletitle{CodeGeeX: {A} Pre-Trained Model for Code Generation with Multilingual Evaluations on HumanEval-X}.
\newblock \bibinfo{journal}{\emph{CoRR}}  \bibinfo{volume}{abs/2303.17568} (\bibinfo{year}{2023}).
\newblock


\bibitem[Zhong et~al\mbox{.}(2009)]%
        {DBLP:conf/ecoop/ZhongXZPM09}
\bibfield{author}{\bibinfo{person}{Hao Zhong}, \bibinfo{person}{Tao Xie}, \bibinfo{person}{Lu Zhang}, \bibinfo{person}{Jian Pei}, {and} \bibinfo{person}{Hong Mei}.} \bibinfo{year}{2009}\natexlab{}.
\newblock \showarticletitle{{MAPO:} Mining and Recommending {API} Usage Patterns}. In \bibinfo{booktitle}{\emph{{ECOOP} 2009 - Object-Oriented Programming, 23rd European Conference, Genoa, Italy, July 6-10, 2009. Proceedings}} \emph{(\bibinfo{series}{Lecture Notes in Computer Science}, Vol.~\bibinfo{volume}{5653})}, \bibfield{editor}{\bibinfo{person}{Sophia Drossopoulou}} (Ed.). \bibinfo{publisher}{Springer}, \bibinfo{pages}{318--343}.
\newblock


\bibitem[Zhou et~al\mbox{.}(2019)]%
        {DBLP:conf/nips/ZhouLSD019}
\bibfield{author}{\bibinfo{person}{Yaqin Zhou}, \bibinfo{person}{Shangqing Liu}, \bibinfo{person}{Jing~Kai Siow}, \bibinfo{person}{Xiaoning Du}, {and} \bibinfo{person}{Yang Liu}.} \bibinfo{year}{2019}\natexlab{}.
\newblock \showarticletitle{Devign: Effective Vulnerability Identification by Learning Comprehensive Program Semantics via Graph Neural Networks}. In \bibinfo{booktitle}{\emph{Advances in Neural Information Processing Systems 32: Annual Conference on Neural Information Processing Systems 2019, NeurIPS 2019, December 8-14, 2019, Vancouver, BC, Canada}}, \bibfield{editor}{\bibinfo{person}{Hanna~M. Wallach}, \bibinfo{person}{Hugo Larochelle}, \bibinfo{person}{Alina Beygelzimer}, \bibinfo{person}{Florence d'Alch{\'{e}}{-}Buc}, \bibinfo{person}{Emily~B. Fox}, {and} \bibinfo{person}{Roman Garnett}} (Eds.). \bibinfo{pages}{10197--10207}.
\newblock


\bibitem[Zhou et~al\mbox{.}(2022)]%
        {DBLP:journals/tse/ZhouYCHMG22}
\bibfield{author}{\bibinfo{person}{Yu Zhou}, \bibinfo{person}{Xinying Yang}, \bibinfo{person}{Taolue Chen}, \bibinfo{person}{Zhiqiu Huang}, \bibinfo{person}{Xiaoxing Ma}, {and} \bibinfo{person}{Harald~C. Gall}.} \bibinfo{year}{2022}\natexlab{}.
\newblock \showarticletitle{Boosting {API} Recommendation With Implicit Feedback}.
\newblock \bibinfo{journal}{\emph{{IEEE} Trans. Software Eng.}} \bibinfo{volume}{48}, \bibinfo{number}{6} (\bibinfo{year}{2022}), \bibinfo{pages}{2157--2172}.
\newblock


\bibitem[Zou et~al\mbox{.}(2021)]%
        {DBLP:journals/tosem/ZouZXLJY21}
\bibfield{author}{\bibinfo{person}{Deqing Zou}, \bibinfo{person}{Yawei Zhu}, \bibinfo{person}{Shouhuai Xu}, \bibinfo{person}{Zhen Li}, \bibinfo{person}{Hai Jin}, {and} \bibinfo{person}{Hengkai Ye}.} \bibinfo{year}{2021}\natexlab{}.
\newblock \showarticletitle{Interpreting Deep Learning-based Vulnerability Detector Predictions Based on Heuristic Searching}.
\newblock \bibinfo{journal}{\emph{{ACM} Trans. Softw. Eng. Methodol.}} \bibinfo{volume}{30}, \bibinfo{number}{2} (\bibinfo{year}{2021}), \bibinfo{pages}{23:1--23:31}.
\newblock


\end{thebibliography}


\end{document}